\documentclass[hyperref]{aa}  
\usepackage{amsmath}
\usepackage{natbib}
\usepackage{graphicx}
\usepackage{subcaption}
\usepackage{float}
\usepackage{url,color,geometry}
\usepackage{txfonts}

\usepackage{wasysym}
\usepackage{mathtools}
\usepackage{longtable, fancyhdr}

\begin{document} 

   \title{Modelling the atmosphere of the carbon-rich Mira RU~Vir \thanks{Based on observations made with ESO telescopes at La Silla Paranal Observatory under program IDs: $085$.D-$0756$ and $093$.D-$0708$}}

   \author{G. Rau \inst{1}
          \and
          C. Paladini \inst{2}
          \and
          J. Hron \inst{1}                    
          \and
          B. Aringer \inst{3}
          \and
          M. A. T. Groenewegen \inst{4}          
          \and
          W. Nowotny \inst{1}}

   \institute{University of Vienna, Department of Astrophysics, T\"urkenschanzstrasse 17, A-1180 Vienna\\
              \email{gioia.rau@univie.ac.at}
         \and
             Institut d’Astronomie et d’Astrophysique, Universite' libre de Bruxelles, Boulevard du Triomphe CP $226$, B-1050 Bruxelles, Belgium
         \and             
             Department of Physics and Astronomy G. Galilei, Vicolo dell'Osservatorio 3, 35122 Padova, Italy
         \and
             Royal Observatory of Belgium, Ringlaan 3,  B-1180 Brussels, Belgium
             }

   \date{Submitted to A\&A. Received April 27, 2015; accepted August 29, 2015.}

 
  \abstract
   {We study the atmosphere of the carbon-rich Mira \object{RU Vir} using the mid-infrared high spatial resolution interferometric observations from VLTI/MIDI.}
   {The aim of this work is to analyse the atmosphere of the carbon-rich Mira RU~Vir, with hydrostatic and dynamic models, in this way deepening the knowledge of the dynamic processes at work in carbon-rich Miras.}
   {We compare spectro-photometric and interferometric measurements of this carbon-rich Mira AGB star, with the predictions of different kinds of modelling approaches (hydrostatic model atmospheres plus MOD-More Of Dusty, self-consistent dynamic model atmospheres). A geometric model fitting tool is used  for a first interpretation of the interferometric data. }
   {The results show that a joint use of different kind of observations (photometry, spectroscopy, interferometry) is essential to shed light on the structure of the atmosphere of a carbon-rich Mira. The dynamic model atmospheres fit well the ISO spectrum in the wavelength range $\lambda = [2.9, 25.0]~\mu$m. Nevertheless, a discrepancy is noticeable both in the SED (visible), and in the interferometric visibilities (shape and level). A possible explanation are intra-/inter-cycle variations in the dynamic model atmospheres as well as in the observations. The presence of a companion star and/or a disk or a decrease of mass loss within the last few hundred years cannot be excluded but these explanations are considered unlikely. 
	}
   {}

 \keywords{instrumentation: high angular resolution --
   techniques: interferometric --
   stars: AGB and post-AGB --
    stars: atmospheres -- 
    stars: circumstellar matter --
     stars: fundamental parameters}

\authorrunning{Rau et al.}
\titlerunning{Modelling the atmosphere of the carbon-rich Mira RU~Vir}

   \maketitle

\section{Introduction}

   The Asymptotic Giant Branch (AGB) is the late evolutionary stage of low-to-intermediate mass stars (typically $\approx 0.8$ to $\approx 8$M$_\odot$). These objects are characterized by a C-O core and He/H burning shells, surrounded by a convective envelope, while the atmosphere consists of atomic and molecular gas, including dust grains. Stellar pulsation can generate shock waves that run through the atmosphere. These shocks can propagate outward, causing a levitation of the outer atmosphere layers, improving the conditions for the formation of dust, which may then lead to a dust-driven wind.

At the early stages of their life, AGB stars have a carbon-to-oxygen-ratio below one. The third dredge-up can turn the chemistry of these objects from oxygen- into carbon-rich \citep{ibenrenzini83}. Carbon-rich AGB stars are important contributors to the enrichment of the interstellar medium, and in their spectra there are signs of carbon-bearing molecules such as C$_2$, C$_3$, C$_2$H$_2$, CN or HCN, while the dust is mainly dominated by amorphous carbon dust grains and SiC \citep{loidl01,yam00}. 

By studying the stellar atmospheres of AGB stars, we hope to better understand many processes that can occur, e.g. the dust formation and mass loss via strong stellar winds, the connection between pulsation and atmospheric structure.
In the case of C-rich AGB stars with no pronounced pulsation, most of the observables derived from hydrostatic models agree fairly well with the observations \citep{aringer09}. However, as the star evolves, time-dependent processes become more important, and the atmosphere expands, causing a decrease of the effective temperature. Hence, model atmospheres that take into account those dynamic processes (pulsation, dust formation, mass loss) are necessary (e.g.\ \citealp{bowen1988, Fleischer92, HoefnerDorfi, hofner03}; see also the excellent reviews from \citealp{woitke03} and \citealp{hofner07}). 

Over the last years there has been an increase in the number of works that investigate the atmosphere of AGB stars combining various techniques \citep{witt01,witt08,witt11,neilson08,martividal11}. On the other hand, to date only few interferometric observations of carbon stars have been compared with model atmospheres \citep{sacuto11,paladini11,cruzal13,txpsc,vanbelle13}.

In this paper we intend to study the atmosphere of the carbon-rich Mira RU Vir, by means of photometry, spectroscopy, interferometry, and model atmospheres \citep{mattsson10, aringer09}, using the most recent grid of dynamic atmosphere models and synthetic spectra \citep{Erik14}. RU~Vir is a C-rich AGB Variable Star of Mira type. It has a period of $433.2$ days in the $V$ band \citep{GCVS}. Its distance is $910$ pc (based on the Period-Luminosity relation from \citealp{whitelock06}), the amplitude of the variability in the $V$ band is $5.2$ mag and the average  value of the magnitude in $V$ is $<V> = 11.6$~mag. RU~Vir is surrounded by a carbon-rich dusty shell made of amorphous carbon (AmC) and silicon carbide (SiC). The latter shows its presence through the strong emission feature around $\sim 11~\mu$m. 

We present MIDI (MID-infrared Interferometric instrument) observations (P.I. G. Rau: ID $093$.D-$0708$(A)) probing regions of dust formation in the circumstellar envelope \citep{walter2,danchi94}. In a forthcoming paper, we will also consider a larger sample of carbon-rich AGB stars. 

In Sect.~\ref{obs} we present the photometry, the spectroscopy, and the interferometric observations of RU~Vir and describe the data reduction performed. Section~\ref{geom} explores the geometry of the environment, and the fit of the geometrical models to the interferometric visibilities to constrain the morphology and the brightness distribution of the object. In general the visibility V as a function of spatial frequencies \textit{u} and \textit{v}, as described e.g. in \cite{winters95}, is the two-dimentional Fourier transform of the intensity distribution $I(\eta,\zeta)$, where $\eta$ and $\zeta$ are the corresponding coordinates on the celestial sphere. In the special case of a circular symmetric intensity distribution, which is produced by spherical symmetric models, the visibility is given by the Hankel transform of the intensity profile: $$V(q) = \int_0^{\phi_{max}} I(\phi)~\phi J_0 (2\pi \phi q )~d\phi~,$$ where $q = \sqrt{u^2 + v^2}$ is the spatial frequency, $\phi$ is the angular separation from the center of the star ($\phi = r/d$) with $d$ the distance of the star, $r$ the radius and $J_0$ the zeroth-order Bessel function. Therefore, the term interferometric visibilities will be from now on referred to as visibilities V, used to denote the normalized quantity $V(q)/V(0)$. In Sect.~\ref{hydr_mod} and \ref{dyn} we investigate the modelling of the atmosphere of RU~Vir, from the hydrostatic and hydrodynamic point of view. In Sect.~\ref{discrep} we propose some of the possible scenarios that could justify the optical excess (Sect.~\ref{optical_excess}) and the discrepancy in the visibilities (Sect.~\ref{visdiscrep}). Moreover, a discussion on the stellar parameters can be found in Sect.~\ref{discussion}. Lastly, we present the conclusions and perspectives for future works, in Sect.~\ref{conclu}.


\section{Observationoal data for RU Vir}\label{obs}

 \subsection{ISO spectrum}\label{isospec}
RU~Vir was observed with the Short Wavelength Spectrometer on board the ISO satellite (SWS, \citealp{isosws}) once on $20$ July 20, 1996 (\citealp{sloan03}). This ISO/SWS spectrum covers the wavelength range form $2.36$ to $45.35~\mu$m, of a spectral resolution of R~$\approx~200$. To the purpose of our investigations, an error of $\pm~10\%$ is assumed for wavelengt $<4.05~\mu$m, and $\pm~5$\% towards redden wavelengths (\citealp{sloan03}). The ISO spectrum can be seen in black in Fig.~\ref{corrected_iso}.

\subsection{Photometry}\label{phot}
RU~Vir photometry in the filters $B$, $V$, $R$, $I$, $J$, $H$, $K$, is given in Table~\ref{table_photometry}. For our study, following \cite{walter2}, we pick the $B$, $V$, $R$, $I$ observations from \cite{eggen1975a} and $J$, $H$, $K$ observations from \cite{whitelock06}. The photometry is taken at the same phase $\phi$ as the ISO/SWS spectrum, which can be calculated in the following way:

\begin{equation} \label{eq:phase}
\phi_{\text{ISO}}^{RU~Vir}  = \frac{(t - T_0) \lvert P\rvert}{P} = 0.65 \pm 0.1
\end{equation}

where $t$ is the time of the observation, in Julian date, $T_0$ is the selected phase-zero point, corresponing to the nearest visual light maximum of the star. The light curves were interpolated to this phase \cite[Fig.$~8$ there]{walter2}. The uncertainties of the different photometric measurements are estimated from the uncertainty in the phase and the scatter of the phased light curve data.

\begin{figure*}
\centering
\includegraphics[angle=180, bb=0 0 792 612, clip=true, scale=0.6]{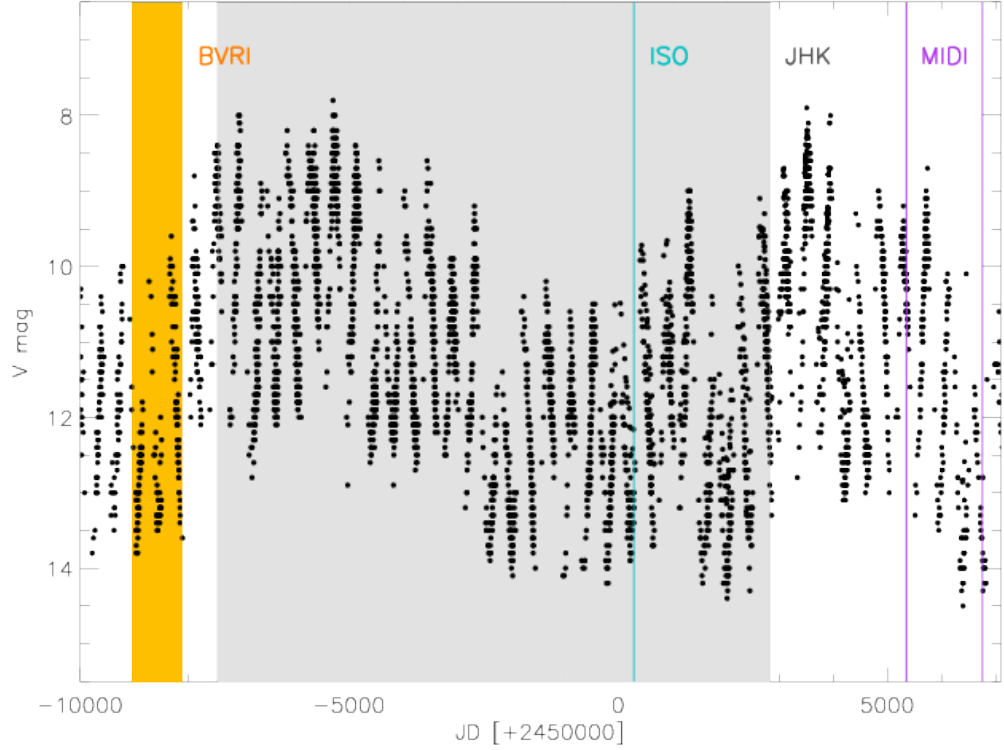}
\caption{AAVSO light curve of RU~Vir in black. The gray shadow shows the time range of the $BVRI$ photometry observed by \cite{eggen1975a}. The vertical green, blue, and violet lines denote the epoch of the ISO spectrum, $JHK$ photometry, MIDI obsetvation respectively.}
\label{newlight}
\end{figure*}

The visual light curve of RU~Vir from AAVSO observations is shown in Fig.\ref{newlight}. The time of the observation of the ISO spectrum and the time interval of \cite{eggen1975a} photometry (that covers more than one pulsation cycle) are underlined in vertical green and blue lines, respectively. The light curve favor the presence of a secondary period on a timescale of $\sim 30$yr \citep{percy99}. 

	
\begin{table*}
\caption{\label{table_photometry} RU~Vir photometry in different filters adopted from \cite{walter2}, in units of mag. The photometry is interpolated to the phase of the ISO$/$SWS spectrum (see Sect.~\ref{phot}).} 
\centering
\begin{tabular}{lllllll}
\hline\hline             
 $B$ & $V$ & $R$ & $I$ & $J$ & $H$ & $K$ \\
\hline
 $15.95 \pm 0.90$ & $11.75 \pm 0.70$ & $8.50 \pm 0.50$ &  $6.90 \pm 0.50$ & $5.50 \pm 0.20 $ & $3.75 \pm 0.15 $ & $2.30 \pm 0.20$ \\
\hline
\end{tabular}
\end{table*}

 \subsection{MIDI data}\label{mididata}
RU~Vir was observed in $2010$ (ID: $085$.D-$0756$(C)) and $2014$ (ID: $093$.D-$0708$(A)) with the $1.8$m Auxiliary Telescopes (ATs) of the Very Large Telescope Interferometer MIDI \citep{midi2003}. MIDI provides wavelength dependent visibilities, photometry, and differential phases in the $N$-band ($\lambda_{\text{range}} = [8,13]$ $\mu$m).
 
\begin{figure}
   \centering
   \includegraphics[width=\hsize, bb=111 382 549 711]{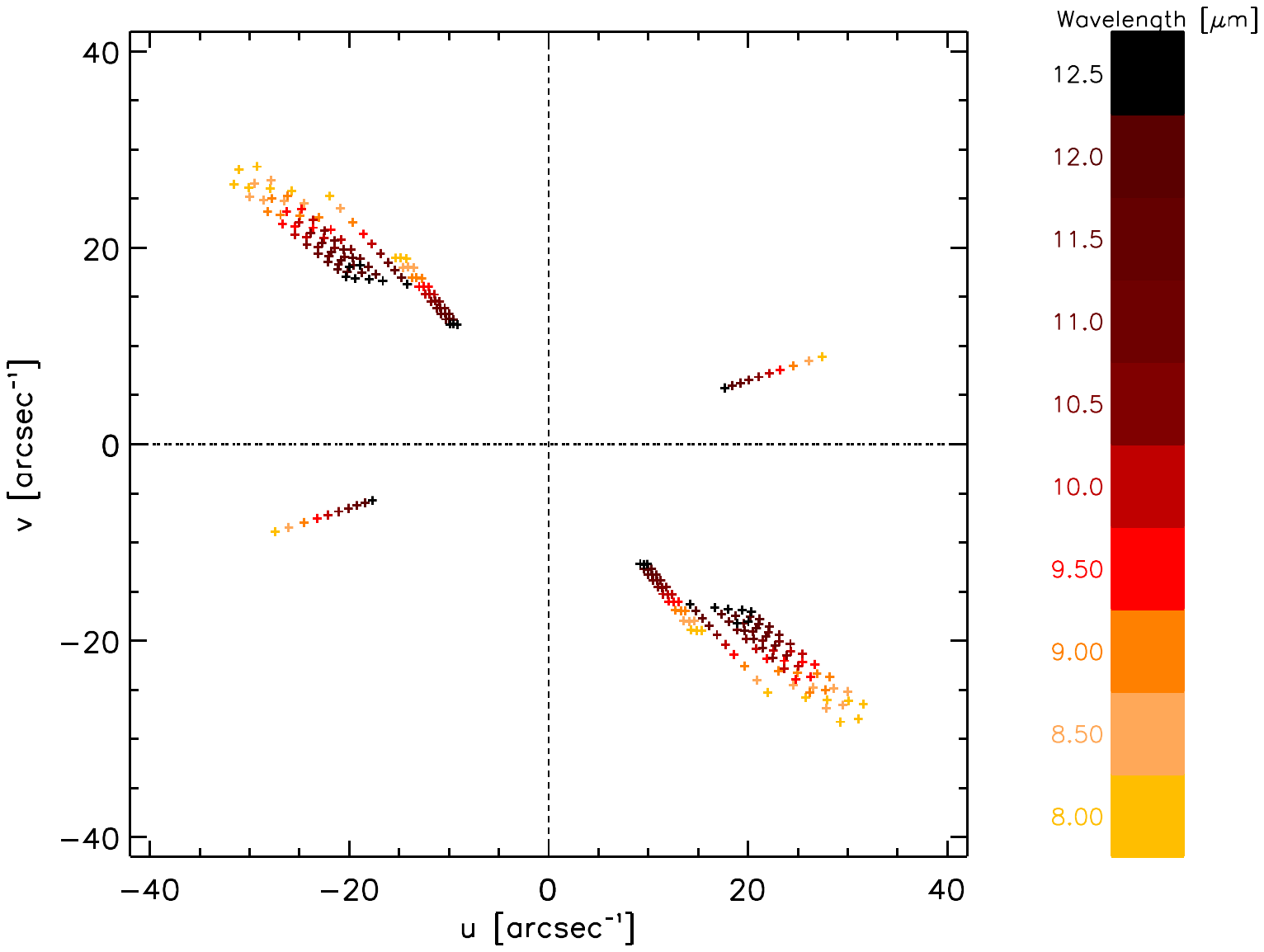}
      \caption{ $uv$-coverage of the MIDI RU~Vir observations listed in Table~\ref{tab_midi_observ}, dispersed in wavelengths.}.
         \label{uv_ruvir}
   \end{figure}

The journal of the observations is available in Table~\ref{tab_midi_observ} of the appendix. The calibrators used are listed below the corresponding science, and the second-last column identifies if the observations are carried out in SCI-PHOT or HIGH-SENS mode. All the observations are in low spectral resolution ($R = 30$). The $uv$-coverage is shown in Fig.~\ref{uv_ruvir}. The main characteristics of the calibrators are listed in Table~\ref{tab_calibrators}. 

The choice of the calibrators follows the selection criteria described in \cite{klotz12}. 
 
\begin{figure}
   \centering
   \includegraphics[angle=90, width=\hsize, bb=72 80 519 686]{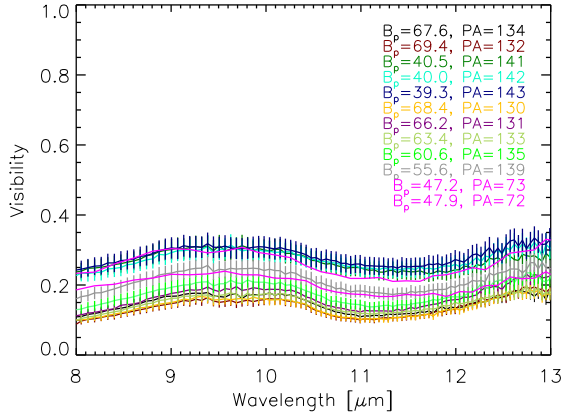}
      \caption{RU~Vir calibrated visibilities, at different baselines and projected angles.}
         \label{ruvir_calibr_vis}
   \end{figure}

Data were reduced with the latest data reduction software package MIA+EWS \citep{jaffe04}. The calibrated visibilities at different baselines are plotted in Fig.~\ref{ruvir_calibr_vis}. The wavelength dependent visibility exhibits the typical shape of the visibilities for carbon stars with dust shells containing SiC grains (Paladini et al. in prep.). There is a drop between $8-9~\mu$m due to C$_2$~H$_2$~+~HCN opacities, and another decline in the visibility shape at $\sim~11.3~\mu$m, due to SiC.

The difference in position angle between the two sets of the observations (in $2010$ and $2014$) does not allow us to perform a check for interferometric variability because they do not probe the same spatial frequencies. So far, interferometric variability was observed only for one star (V~Oph, \citealp{ohnaka07}); therefore we decided to analyze the data all at the same time. When it comes to study the geometry (Sect. \ref{geom}) we will keep in mind that if these two points push for an asymmetric solution, then variability could be still an explanation.
 
\begin{figure}
\centering
\includegraphics[angle=90, width=\hsize, bb=0 0 504 684]{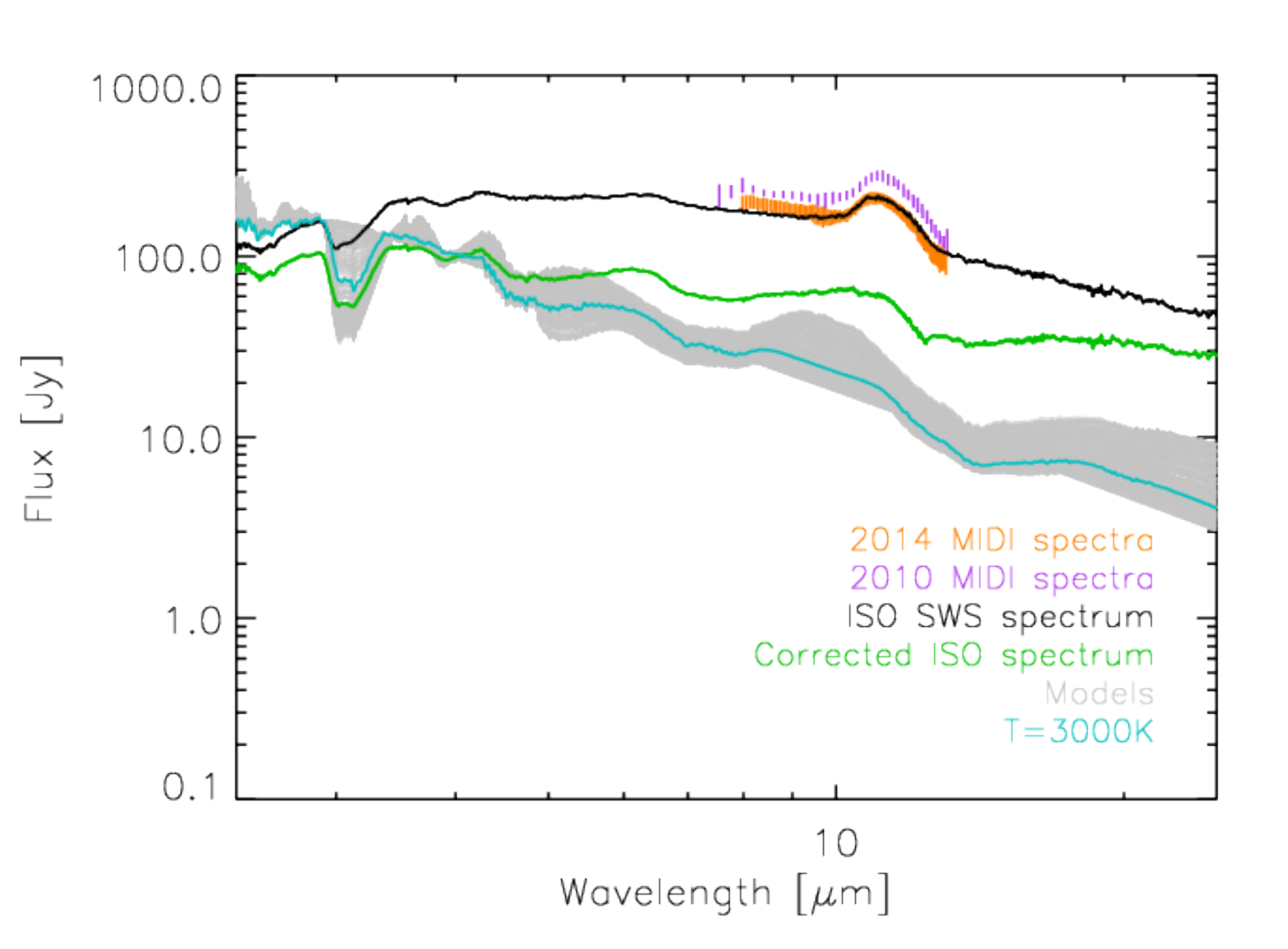}
\caption{Comparison of observational data with modelling results. The black line shows the ISO/SWS spectrum of  RU~Vir, the green line the SED corrected from dust emission. The orange and violet shadows marks the MIDI spectra, while the gray one denote hydrostatic models grid. The best fitting model is shown in cyan.}
\label{corrected_iso}
\end{figure}

The MIDI differential phase is zero, and the MIDI spectra are shown in Fig.~\ref{corrected_iso}. The shape of the MIDI spectra agree within the errors  with the ISO spectrum. The difference in flux level between the two sets of MIDI observations is typical for C-rich Miras \citep{hron98}.

\begin{table}
\caption{\label{tab_calibrators} Parameters of the calibrator targets.}
\centering
\begin{tabular}{llll}
\hline\hline             
Target & Spectral type & F$_{12}$~\textsuperscript{a} & Diameter~\textsuperscript{b} \\
 &  & [Jy] & [mas] \\
\hline

HD$120323$ & M4.5III  & $255.4$ & $13.25 \pm 0.060$  \\
HD$133216$ & M3/M4III & $200.7$ & $11.154 \pm 0.046$  \\        
HD$81797$  & K3II-III & $157.6$ & $9.142 \pm 0.045$    \\
\hline
\end{tabular}
\tablefoot{(a)IRAS Point Source Catalog: \url{http://simbad.u-strasbg.fr/simbad/}.\newline 
(b)\url{http://www.ster.kuleuven.ac.be/~tijl/MIDI_calibration}}
\end{table}

\section{Geometrical models}\label{geom}

   \subsection{Geometry of the environment}
Before comparing the model atmospheres and the data, we studied the morphology of the circumstellar environment, interpreting the MIDI interferometric data with geometric models.

This has been done by using the geometrical model fitting tool GEM-FIND (GEometrical Model Fitting for INterferometric Data) of \citealp{gemfind}. The program fits geometrical models to wavelength dependent visibilities in the $N$-band, to constrain the morphology and brightness distribution of an object. A detailed description of the fitting strategy and of the $\chi^2$ minimization procedure can be found in \cite{klotz12}.\newline

Different models with optically thick (models with one-component) or thin (models with two-components) circumstellar envelopes were tested. In Table~\ref{table_gemfind_results} we list the corresponding reduced $\chi^2$ values, and the parameters fitting best. In the case of elliptic models the ratio of major and minor axis is shown. For the two component models with circular uniform disc (UD)~+~circular gaussian, the brightness ratio is indicated. In the latter case the fit was performed with a fixed value of the angular diameter, to avoid to have too many free parameters in the fit. We started the fit of the composite model imposing the diameter to be equal to the one derived from the $\theta/(V-K)$ relation of \cite{vanbelle13}: $\theta = 2.18~$mas. Then we increased the diameter by steps of $2$~mas until a $\theta_{\rm{max}} = $20~mas.  

One component spherically symmetric and elliptical models have large reduced $\chi^2$ and are not able to reproduce the data. A better fit to the data is obtained by a two-components model, including a circular UD plus a gaussian distribution of the circular dusty envelope. This yields a reduced $\chi^2_{\text{min}}$ of $0.96$ for a central star simulated by a UD with angular diameter at $11~\mu$m of $18~$mas (see Table~\ref{table_gemfind_results}), probably because of molecular opacities. 

\begin{figure}
   \centering
   \includegraphics[angle=90, width=\hsize, bb=318 79 554 702]{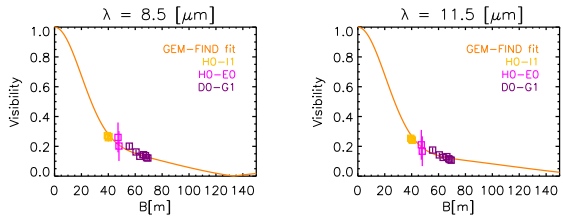}
      \caption{Comparison of observational data with the best fitting geometrical model. The calibrated visibilities are shown versus baseline length, for two different wavelengths (left: 8.5~$\mu$m; right: 11.5~$\mu$m). The symbols represent MIDI observations at three different baseline configurations. The lines show the best-fitting model consisting of a circular UD and a circular gaussian.}
         \label{fit_gemfind_base-vis}
   \end{figure}

In Fig.~\ref{ruvir_calibr_vis} we show the the visibilities vs. wavelength, while in Fig.~\ref{fit_gemfind_base-vis} the fit of the visibility vs. baseline at different wavelength is shown. 

In general, the geometric models show that, for the u-v covered by our RU~Vir observations, there is no major asymmetry detected.

 \begin{table*}
\centering
\caption{\label{table_gemfind_results} GEM-FIND results. For each fit (one or two components), the full width at half maximum and angular diameter at $11~\mu$m are given. When applicable also the ratio of minor and major axis and the brightness ratio are given. The  best fitting model is shown in bold face.}
\begin{tabular}{llllll}
\hline
\hline
Geometric model  & \textbf{$\chi^2_{\text{red}}$}  & FWHM$_{11}$    &  $\theta_{11}$  &  a/b &  B$_{\text{ratio}}$  \\
                   &   &    [mas]          &      [mas]  &          &     \\     
\hline
\hline
One component     &      &               &                     &              &             \\
\hline
Circular UD       & $40.10$ &  $\cdots$          &  $38.10 \pm 0.17$   &   $\cdots$   &   $\cdots$  \\
Circular Gauss    & $15.42$ &  $28.44 \pm 0.24$  &  $\cdots$           &   $\cdots$   &   $\cdots$  \\
Elliptic UD       & $12.31$ &  $\cdots$          &  $177.34 \pm 0.77$  &   $0.2$      &   $\cdots$  \\
Elliptic Gauss    & $5.41$  &  $\cdots$          &  $56.62 \pm 0.44$   &   $0.3$     &    $\cdots$   \\
\hline
Two components     &        &             &                     &              &             \\
\hline
Circ UD+Circ Gauss  & $1.59$ & $37.88 \pm 0.99$  &  $2.81$~\textsuperscript{a} &$\cdots$ & $0.12 \pm 0.01$\\
Circ UD+Circ Gauss   & $1.57$ & $38.02 \pm 1.01$  &  $6.00$~\textsuperscript{a} &$\cdots$ & $0.12 \pm 0.01$\\
Circ UD+Circ Gauss   & $1.53$ & $38.18 \pm 1.02$  &  $8.00$~\textsuperscript{a} &$\cdots$ & $0.13 \pm 0.01$\\
Circ UD+Circ Gauss   & $1.49$ & $38.38 \pm 1.05$  &  $10.00$~\textsuperscript{a}&$\cdots$ & $0.13 \pm 0.01$\\
Circ UD+Circ Gauss   & $1.43$ & $38.65 \pm 1.08$  &  $12.00$~\textsuperscript{a}&$\cdots$ & $0.15 \pm 0.01$\\
Circ UD+Circ Gauss   & $1.35$ & $39.01 \pm 1.12$  &  $14.00$~\textsuperscript{a}&$\cdots$ & $0.15 \pm 0.01$\\
Circ UD+Circ Gauss   & $1.27$ & $39.47 \pm 1.18$  &  $16.00$~\textsuperscript{a}&$\cdots$ & $0.17 \pm 0.01$\\
Circ UD+Circ Gauss    & $1.15$  & $40.08 \pm 1.12$  &  $18.00$~\textsuperscript{a}&$\cdots$& $0.19 \pm 0.01$\\
\textbf{Circ UD+Circ Gauss}  & $1.01$ & $40.90 \pm 1.38$  &  $20.00$~\textsuperscript{a}&$\cdots$& $0.21 \pm 0.01$\\
\hline
\end{tabular}\\
\textbf{Notes}. (a): Models with central star diameter fixed at the values indicated above during the fit. 
\end{table*}

\section{Hydrostatic model atmospheres + MOD models of dusty envelopes}\label{hydr_mod}
Following \cite{sacuto11}, we perform the first step of our study first by fitting the ISO spectrum with hydrostatic model atmospheres (COMARCS models, \citealp{aringer09}). 

Each hydrostatic model is identified by the following initial parameters: mass, effective temperature, C/O, log(g), metallicity. Details about these model atmospheres and the computed spectra are described in \cite{aringer09}. The resolution of each synthetic spectrum\footnote{Available at \url{http://stev.oapd.inaf.it/synphot/Cstars/}} is R$~=~200$ to match the resolution of ISO, and the spectral range $[0.4, 45.0]~\mu$m.

Since RU~Vir is a dust-enshrouded carbon-Mira, we included dust by using the radiative transfer code MOD (More Of Dusty, \citealp{mod}). This code is based on DUSTY, that is a publicly available $1$\,D dust radiative transfer code (\citealp{dustypaper}), whose mathematical formulation is described in detail in \cite{ivezic97}. MOD tries to optimize a set of parameters (e.g. luminosity, dust optical depth, dust condensation temperature and slope of the density law) with some constraints as spectro- photometric-data, $1$D intensity profiles and visibility curves. A quantitative measure of the quality of the fit is obtained through performing a $\chi^2$ minimization, where the $\chi^2$ are computed separately for all the four types of observations. We note at this point that the dust condensation temperature in this context actually is the dust temperature at the inner boundary of the dust shell. Following previous work (e.g.\ \citealp{mod}) we will call it the dust condensation temperature.

Following \cite{sacuto11}, we use a mixture of $90~\%$ AmC and $10~\%$ SiC for the dust composition. This ratio gave the best results for the ISO/SWS spectrum of the carbon star R~Scl. Fits for RU~Vir with a lower SiC fraction were not as good. The optical constants for opacities of silicon carbide and amorphous carbon are taken from \cite{pitman08} and \cite{roleau_and_martin} respectively. As the standard grain size distribution of DUSTY produced much too peaked SiC features, we used the Distribution of Hollow Spheres (DHS, \citealp{mod} and references therein) with a mean grain size of the dust mixture of 0.2~$\mu$m. This grain size is similar to the typical sizes found from models for dust driven mass loss \citep{mattsson11}. However, we note that these models include only amorphous carbon grains. On the other hand, SiC grains found in presolar meteorites have sizes of the order of 1~$\mu$m (e.g.\ \citealp{gail}), indicating that our mean size is consistent with laboratory data. The chosen grain size distribution then, gave a good fit of the SiC feature.

\begin{table*}
\caption{Output values of MOD fitting, for three different cases. The parameters are: luminosity L, optical depth $\tau$, dust condensation temperature T$_c$, parameter of the slope of the density law p. The last four lines show the reduced $\chi^2$ for the fit of: photometry, spectroscopy, interferometric visibilities, and then the total one.}
\label{tab_output_mod} 
\centering
\begin{tabular}{llll}
\hline\hline
$ $  & T$_{\text{c}}^{\ast} = 1000$~K  &    T$_{\text{c}}^{\ast} = 1300$~K   &   all free   \\
\hline\hline	
L~[$L_\odot$] & $6118.05 \pm 32.54$  & $5770 \pm 28.02$ &   $6322.66 \pm 51.07$ \\
$\tau$          & $7.96 \pm 0.10$  &  $12.52 \pm 0.15$  &   $7.11 \pm 0.24$ \\
T$_{\text{c}}$~[K]   & $1000$   &  $1300$   &   $1008.68 \pm 12.27 $  \\	
p   &  $2.54 \pm 0.01$     &  $2.19 \pm 0.01$ &   $2.28 \pm 0.01$ \\
$\chi^2_{\text{red}}$[pho]  & $22.41$   & $63.31$  &  $9.86$ \\
$\chi^2_{\text{red}}$[spe]  & $3.35$  &   $3.66$ &   $5.63$\\
$\chi^2_{\text{red}}$[int]  & $60.13$ & $161.77$ &   $65.88$\\
$\chi^2_{\text{red}}$[TOT]  & $6.12$ &  $11.56$ &   $7.90$\\
\hline
\end{tabular}\\
\textbf{Notes}. Parameters without errors were fixed (not fitted) in MOD. See Sect~\ref{hydr_mod} for details.
\end{table*}

\subsection{The fitting procedure}\label{MOD_fitting_procedure}
The hydrostatic models alone are not able to reproduce the ISO spectrum. This is because the hydrostatic models do not include dust, while the dust content in the atmosphere of RU~Vir is quite prominent, as the spectral energy distribution from ISO/SWS shows (see Fig.~\ref{corrected_iso}, black line). 
In the following we will use the hydrostatic model atmosphere as inner radiation source for MOD. 
The model parameters are selected as described below.
Considering that RU~Vir is located in the vicinity of the sun, a first selection of the hydrostatic models was made choosing only models with solar metallicity. To the aim of being independent of the distance each hydrostatic model spectrum was normalized to the ISO flux at $2.9~\mu$m. We chose this wavelength because here resides the local minimum of the molecular absorption (\citealp{aringer09}). 

\begin{figure}
\centering
\includegraphics[angle=90, width=\hsize, bb=72 67 519 681]{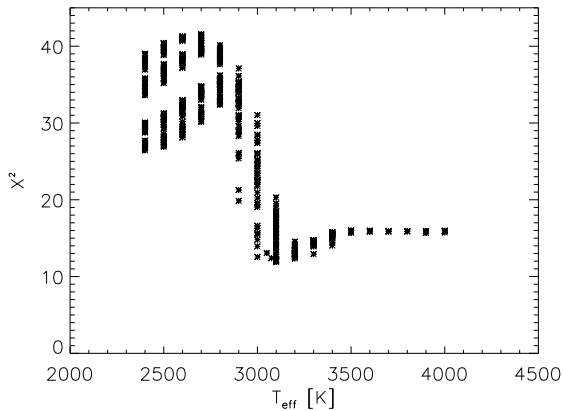}
\caption{Results of the $\chi^2$ fitting for the effective temperature, of the the synthetic spectra based on COMARCS models, with the ISO spectrum. The fit is done in the range of: $[2.9, 3.6]~\mu$m.}
\label{chi1}
\end{figure}

We constrained the effective temperature of RU~Vir by fitting the ISO spectrum with the synthetic spectra, and performing a $\chi^2$ minimization, in the range $[2.9, 3.6]~\mu$m. The molecular features in this wavelength range are good temperature indicators for hydrostatic C-stars \citep{jorgensen00} and provide a first approximation for our case. 
The result obtained in this way is $T = 3100 \pm 100$~K. The C/O is set to C/O$ = 2.0$ because this value is the one corresponding to the lowest $\chi^2$ in the wavelength region above. 
Figure~\ref{chi1} illustrates the $\chi^2$ vs. $T_{\text{eff}}$ in the wavelength range chosen as the best for the parameter determination; the best fitting value is located at the minimum of the $\chi^2$.



We proceed including the dust component with MOD a posteriori, in the way described as follows. We fit with MOD the photometry (see Sect.~\ref{phot}), spectroscopy (see Sect.~\ref{isospec}) and interferometry (see Sect.~\ref{mididata}) of the star. The input parameters of MOD are effective temperature (determined with the hydrostatic model atmosphere fit), distance and the dust properties. The optical depth $\tau$, the parameter that governs the shape of the density law p, and the dust condensation temperature T$_{\text{c}}$ can be fitted or set to a fixed value. 


To assess how much the spectrum used for the inner radiation source is affected by dust emission, an iterative procedure was used. We determined a ''corrected'' ISO spectrum by performing a MOD fit using the 3100\,K spectrum (see Fig.~\ref{corrected_iso}) derived above. We then subtracted the contribution of the dust emission at each wavelength from the total flux, obtaining in this way a stellar photospheric contribution. Repeating the effective temperature determination, results in a best model atmosphere with an effective temperature of 3000\,K, i.e.\ only slightly cooler than the first estimate. This model spectrum was then used for the remaining calculations. We note that the corrected ISO spectrum has its maximum flux at a longer wavelength than the hydrostatic model but the depth of the 3~$\mu$m feature is very similar to the model. This indicates that any difference between the corrected and the original ISO spectrum is caused by residual dust emission and not an incorrect model temperature.


As there is no generally agreed value of the dust condensation temperature in the literature, we performed the fit using three different T$_c$ values: (i) T$_{\text{c}}= 1000~K$, based on the theoretical equilibrium condensation temperature from \cite{gail}, and (ii) T$_{\text{c}}= 1300~K$ (an average of the dust condensation temperature values from \citealp{walter2} and from \citealp{lobel99}). (iii) Finally we allowed all the input parameters to vary freely to verify the behavior of the models. The results are shown in Table~\ref{tab_output_mod}. 

To check the importance of the ISO spectrum for the final result of the MOD fit we made
some experiments using a coarser sampled ISO spectrum and artificially increasing the error associated with the ISO spectrum. These tests show that the full ISO spectrum is essential for constraining the parameters
of the dust envelope. Nevertheless, the real uncertainties of the dust parameters are considerably higher
than their formal errors.

In Fig.~\ref{outputmod} we present the comparison between the photometry (violet circles), the spectroscopy (blue line) and the SED of the best fitting model. We note that the photometry shortward of 1~$\mu$m cannot be well reproduced even if a maximum of free parameters is allowed. An hotter effective temperature cannot solve this discrepancy as it would be incompatible with the strength of the molecular features in the ISO spectrum. Figure~\ref{mod__visib} shows the comparison between MOD visibilities (lines) and observed visibilities (circles). As the observed visibilities do not depend strongly on the wavelength we selected only the two wavelengths corresponding to the maximum and minimum visibility (Fig.~\ref{ruvir_calibr_vis}).
Comparing the spatial frequencies at a given visibility level, one can conclude that the model is roughly a factor $2$ too small. The visibilities could be brought to agreement by assuming a distance of $\sim 300~$pc but this is unrealistically small. Therefore we have to conclude that the model is not extended enough or is missing a component. Another possibility is a difference between the observed and synthetic brightness contrast of multiple shells.

{We will now compare the observations with the DMAs, which in principle would provide a more complete and physically consistent description of RU~Vir.



\begin{figure}
   \centering   
   \includegraphics[angle=90, width=\hsize, bb=74 94 520 696]{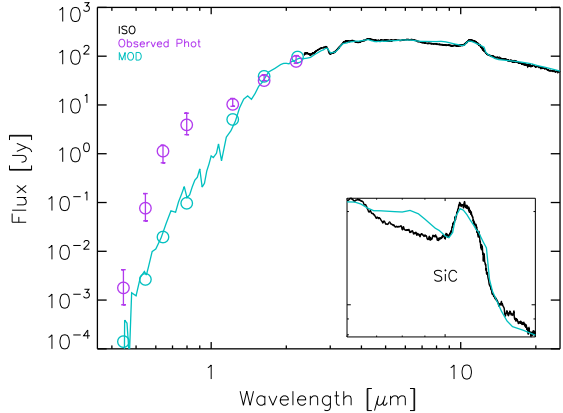}
    \caption{Comparison of observed photometry (Table~\ref{phot})+spectroscopy with the best fitting synthetic spectrum based on hydrostatic model atmospheres and a dusty envelope (see Table~\ref{tab_output_mod} + Sect.~\ref{fitting_procedure}).}
\label{outputmod}
\end{figure}


\begin{figure}
   \centering   
   \includegraphics[angle=90, width=\hsize, bb=74 83 515 681]{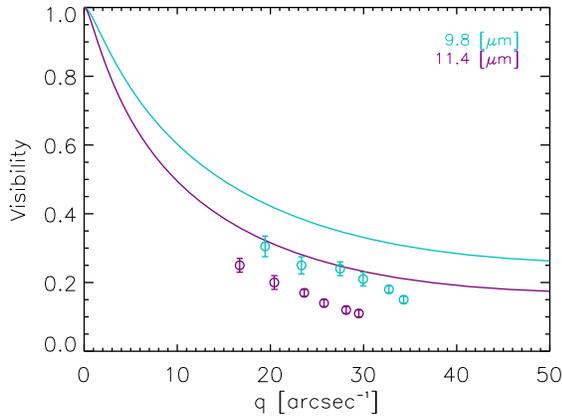}
    \caption{Comparison of MOD visibilities (lines) compared with the VLTI/MIDI observations of RU~Vir (circles), at $9.8~\mu$m (blue) and $11.4~\mu$m (violet).}
\label{mod__visib}
\end{figure}

\section{Dynamic model atmospheres}\label{dyn}
In this section we present the analysis made by using the Dynamic Model Atmospheres (from now on: DMA) from \citet{mattsson10} and model spectra from \citet{Erik14}. For a detailed description of the modelling approach see \citet{HoefnerDorfi}, \citet{hofner99},  \citet{hofner03}, \citet{mattsson10} and \citet{Erik14}). For applications to observations we refer to \citet{loidl99}, \citet{Gautschy-Loidl04}, \citet{nowotny10} and \citet{walter2}. Those models are the result of solving the system of equations for hydrodynamics, frequency-dependent and spherically symmetric radiative transfer, plus a set of equations that describe the time-dependent dust grains formation, growth, and evaporation. The dynamic model start from an initial hydrostatic structure. The stellar pulsation is introduced by a ''piston'', i.e. a variable inner boundary below the stellar photosphere, while the dust formation (only amorphous carbon) is evaluated by the ''method of moments'' \citep{gauger90,gail88}. 
 
The main parameters that characterize the DMA are: luminosity L, effective temperature T$\text{eff}$ (corresponding Rosseland radius, defined by the distance, from the center of the star, to the layer at which the Rosseland optical depth equals unity), [Fe/H], C/O, as well as for the dynamical aspects period P, piston velocity amplitude $\Delta_{\text{u}}$ and the parameter f$_l$ used to adjust the luminosity amplitude of the model. The resulting proprieties of the hydrodynamic calculations are the mean degree of condensation, wind velocity and the mass-loss rate. Each model provides a set of ''time-steps'', representing the different phases of the stellar pulsation. 
 
We computed synthetic spectra using the COMA code and the associated radiative transfer \citep{aringer00,aringer09}. The abundances of all the relevant atomic, molecular and dust species were computed, starting from temperature-density structure, and considering the equilibrium for ionization and molecule formation. The continuous gas opacity as well as the intensities of atomic and molecular spectral lines are subsequently calculated assuming LTE. The corresponding data are consistent with the ones used for the construction of the models and are listed in \cite{cristallo07} and \cite{aringer09}. 

SiC is added a posteriori with COMA by dividing the condensed material from the model into $90$\% amorphous carbon using data from \citep{roleau_and_martin} and $10$\% silicon carbide based on \citep{pegourie}.

In order to be consistent with the model spectra from \citep{Erik14}, we have treated all grain opacities in small
particle limit\footnote{An inconsistent treatment of grain opacities
causes larger errors in the results than neglecting
the grain sizes.} (SPL). The temperature of the SiC particles was assumed to be equal to the one of amorphous carbon. This is justified, since the overall distribution of the absorption is - except for the SiC feature around $11.4~\mu$m - quite similar for both species. As a consequence, the addition of SiC would also not cause significant changes of the thermal model structure. It should be noted that the effects of scattering are not included, since the SPL is adopted.

\subsection{The DMA fitting procedure}\label{fitting_procedure}

Given the difficulties of MOD to reproduce the photometry, and the fact that the ISO/SWS spectrum provides the strongest constraints for gas {\em and} dust, we decided to first base the comparison with DMA only on the ISO/SWS spectrum. Thus we performed a $\chi^2$ minimization in the wavelength range: $2.9 - 25~\mu$m between spectroscopic observation (ISO/SWS spectrum) and each of the $540$ models of the grid (with a total of $\sim~140000$~time steps).
Based on the $\chi^2$ we obtain a best fitting time-step for each model. 

In Fig.~\ref{chisq} the $\chi^2$ vs. average outflow velocity at the outer boundary and vs. mass loss are shown. 
In Table \ref{tab_param_dyn} we list the parameters of the best three time steps of the whole grid and the corresponding $\chi^2$. These time steps cover to the $68$\% confidence region of $\chi^2$. The relevant SEDs, without SiC added, are shown in the left panel of Fig.~\ref{final_plot}. The right panel shows in gray all the time-steps of the model that belongs to the best fitting time-step ($\chi^2 = 9.79$). In pink this best fitting time-step is marked, now with the artificial inclusion of SiC (see also inset in Fig.~\ref{final_plot}). 
 
There is a good agreement between models and observations in the spectral range of the ISO spectrum. 
Even based on this broad range of $\sim~140000$ time-steps, it is not possible to reproduce the photometry shortward of 1$\mu$m. Possible explanations for this disagreement are discussed in Sect.~\ref{discrep}.

   \begin{figure}
   \centering
   \includegraphics[angle=90, width=\hsize, bb=64 55 525 722]{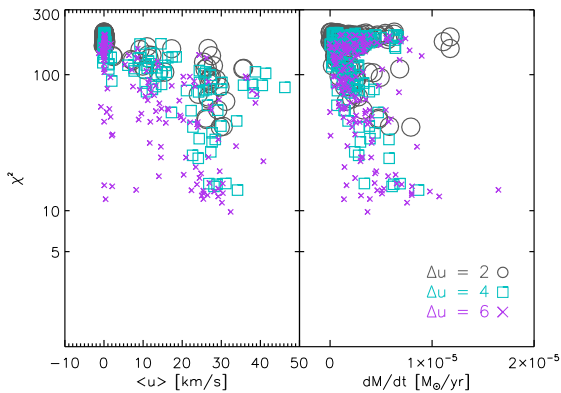}
      \caption{Reduced $\chi^2_{\text{red}}$ vs. average outflow velocity at the outer boundary and vs. mass loss. Different symbols and colors are used for different piston velocities: $2$, $4$, $6$.}
         \label{chisq}
   \end{figure}

   \begin{figure*}
   \centering
   \includegraphics[angle=90, width=\hsize, scale=0.6, bb=0 0 504 684]{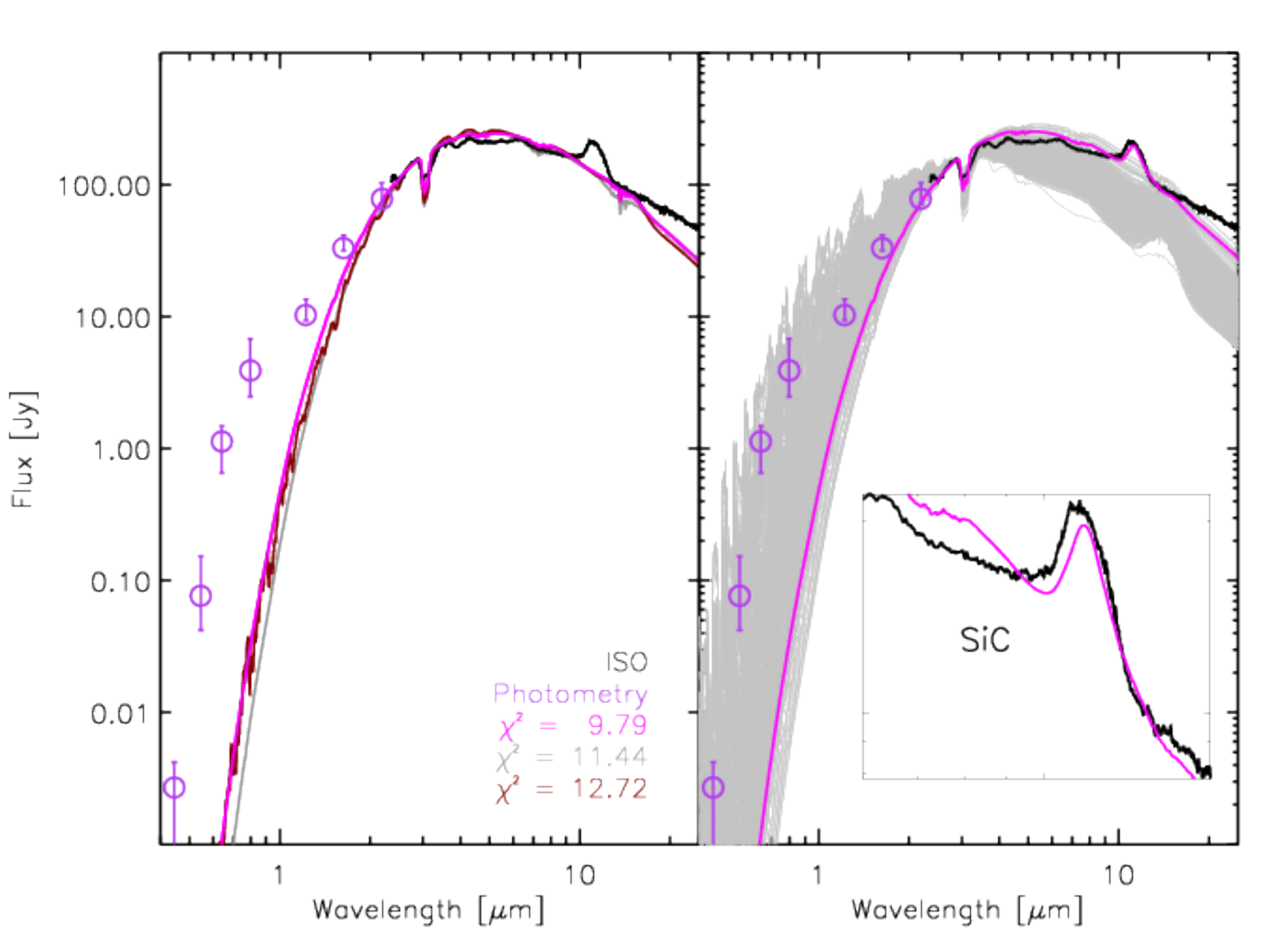}
      \caption{Left: Observational data (ISO spectrum in black line and photometry in violet circles), compared with synthetic spectra of the three best fitting time-steps (belonging to different DMA). Right: Best fitting time-step (pink) compared with ISO (black). The gray lines are the various other phases (time-steps) of the same model.}
         \label{final_plot}
   \end{figure*}


\begin{table*}
\centering
\caption{\label{tab_param_dyn}  Three best fitting time-steps dynamic models from the whole grid from \cite{Erik14} resulting from comparison in Sect.~\ref{midi_dma}. Listed are the corresponding values of the $\chi^2$ for the best fitting time-steps, the parameters, and the phase $\phi$.}
\begin{tabular}{lllllllllll}
\hline
\hline
 $T_{\text{eff}} $ & $lg~L_{\star} $ & $M $ & $log~g$ & $C/O$ & $\Delta u_p$ & $f_L$ & $\dot{M}$&  $\phi$ &  ${\chi^2}_{\text{red}[0.4-25]}$    \\
 $[K]$ & $[L_{\odot}]$ & $[M_{\odot}]$ &   &   &    &  & $[M_{\odot}/yr]$ &    &    \\
\hline
2800&	4.00&	1.5&    -0.66& 2.38&	6&	2&  3.68E-06& 6.50& 9.79   \\
2800&	4.00&	1&	    -0.84& 2.38&	6&	2&  7.56E-06& 0.53& 11.44 \\
2600&	4.00&	1& 	    -0.96& 2.38&	6&	2&  1.44E-05& 3.40& 12.72 \\
\hline
\end{tabular}\\
\end{table*}

\subsection{MIDI data vs. dynamic models}\label{midi_dma}

Following \cite{paladini09} we calculated the intensity and visibility profiles for all the time-steps of the best fitting DMA derived in Sect.~\ref{fitting_procedure}.
 We combined all the interferometric data (even though they are taken at different epochs, see Sect.~\ref{mididata}), and we performed a $\chi^2$ test to obtain the best-fitting visibility profile. We obtained a new best fitting time-step for the visibilities observations of RU~Vir, different from the one fitting best the ISO spectrum 

The time-step best fitting the visibility is shown in Fig.~\ref{intensity_and_vis_dma}. The upper panels illustrate the behavior of the intensity profiles at two wavelengths ($8.5$ and $11.4$~$\mu$m), while the lower panels show the visibility vs. spatial frequencies.
The phase of the best fitting time step is $0.24$, i.e.\ slightly different from that one best fitting the SED (see Sect.~\ref{discrep}).
The best fitting time-step has a Rosseland radius of 2.561 mas (500.9~$R_{\odot}$) and a luminosity of 3982.8~$L_{\odot}$.

The intensity are characterized by a first lobe (the photosphere with the hotter molecular gas) followed by two shells.
The first shell is brighter at $11.4$~$\mu$m, where SiC contributes. Moreover, while at short wavelengths (lower left panel) the visibility can reproduce the data, at long wavelengths (lower right panel) the model is off.

The size of the model stays the same ($\sim~20$~AU), but the flux ratio between the different components is changing.

The discrepancy between the model and the data is also observed when plotting the wavelength dependent visibility for the three baseline configurations (Fig.~\ref{dispersed_vis}).
While at short wavelengths ($8-9$~$\mu$m) the synthetic visibilities match the level of the observations, at longer wavelengths there is a slope in the model that is not observed in the data. The differences between model and observations are more and more pronounced the shorter the baselines is.

    \begin{figure*}
   \centering
   \includegraphics[angle=90, width=\hsize, bb=66 79 554 703]{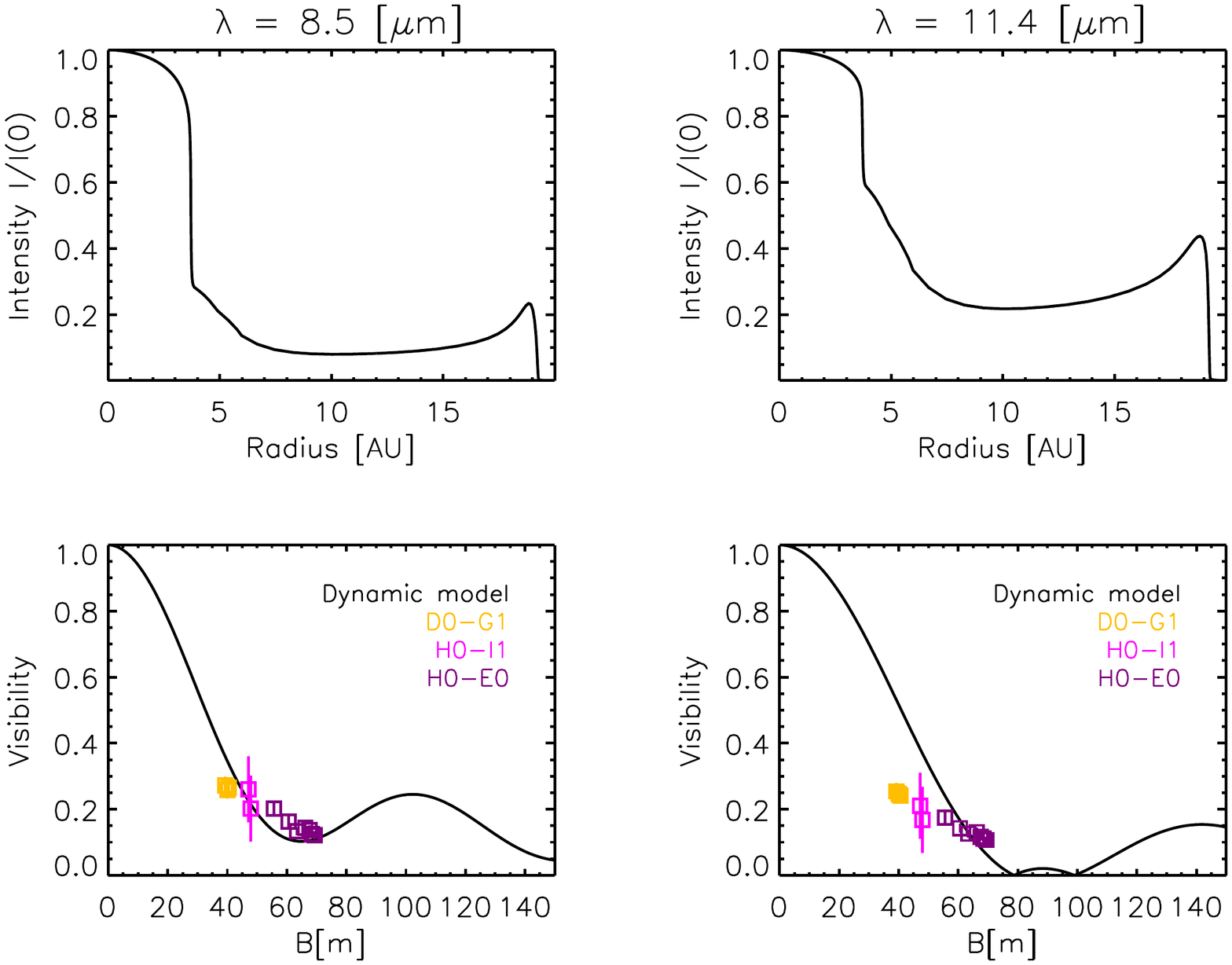}
      \caption{Comparison of interferometric observational data for RU~Vir with the modelling results based on the DMA best-fitting the visibilities. \textbf{Up}: intensity profile at two different wavelengths: $8.5~\mu$m and $11.4~\mu$m. \textbf{Down}: visibility vs. baseline; the black line shows the dynamic model, the colored symbols illustrate the MIDI measurements at different baselines configurations.}
         \label{intensity_and_vis_dma}
   \end{figure*}

\begin{figure*}
\centering
\includegraphics[angle=90, width=\hsize,  bb=66 79 554 703]{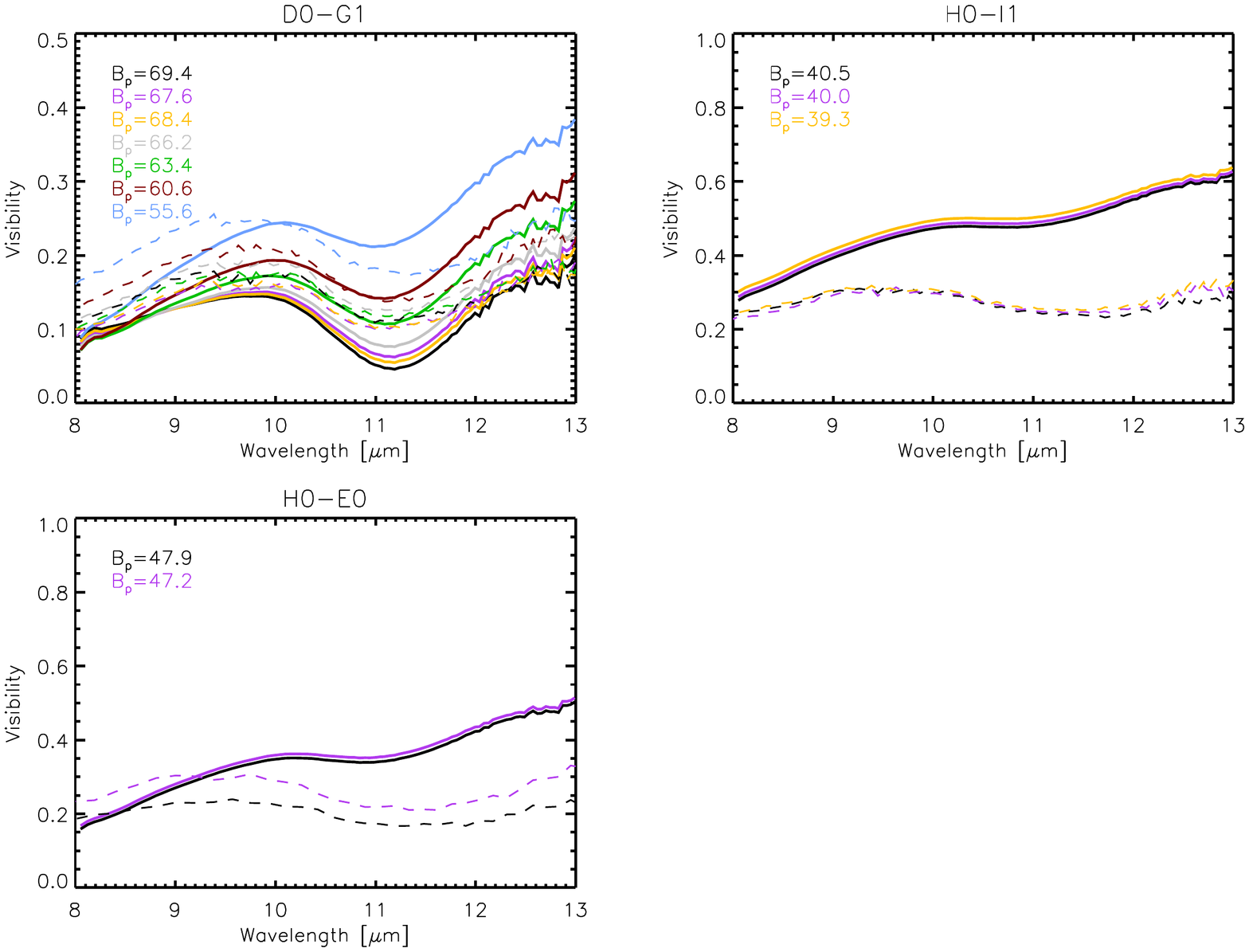}
\caption{Wavelength dependent visibilities in the MIDI range. The different panels show the three different baseline configurations of our observations. Models are plotted in full line, observation in dashed lines, at different projected baselines (see color legend).}
\label{dispersed_vis}
\end{figure*}

\section{Discussion}\label{discrep}

While all our attempts to reproduce the SED (ISO spectrum~+~photometry) and interferometric MIDI data by different models showed some agreement, three major discrepancies remain: 
(1)~No model can describe \textit{both} the optical part and the infrared part of the SED. This may be interpreted either as an excess in the optical or beyond 2~$\mu$m.
(2)~The model visibilities show higher values and/or a different behavior as a function of wavelength than the MIDI data. 
(3)~The DMA predict too low fluxes beyond 14~$\mu$m. 

Since (1) and (2) apply to all model approaches we believe that this is mostly caused by properties of the object which can neither be described by the rather flexible MOD approach nor by the physically consistent DMA (see Sect.~\ref{optical_excess} and \ref{visdiscrep}). In the following we will compare our results with previous work and discuss possible reasons for the discrepancies.

 \subsection{Stellar parameters vs. values from the literature}\label{discussion}
As described in Sect.~\ref{hydr_mod} and \ref{dyn}, stellar parameters were derived from the fit of synthetic spectra based on COMARCS hydrostatic models + MOD to the observations (see Fig.~\ref{outputmod}), and from the study of the SED and MIDI data with the DMAs. 

The best fitting \textbf{optical depth} value  resulting from this work is $\tau_{0.55} = 7.96 \pm 0.10$ at $\lambda = 0.55~\mu$m. \cite{lorenz01} found $\tau = 2.5$ at $1~\mu$m. \cite{dustypaper} indicate an optical depth value at the wavelength $\lambda = 2.2~\mu$m of $\tau_{2.2} = 0.47$. 
Assuming $\tau_{\lambda} \propto 1 / {\lambda}$ and scaling the above literature results to $0.55~\mu$ we arrive at 4.5 and 1.9, respectively. Our value is significantly higher but our result is based on a broader and much better sampled SED which should explain the differences. 

The \textbf{Rosseland radius} of the best fitting time step(s) of $\sim2.6$~mas is close to the ones derived from the diameter $(V-K)$ relations of \cite{vanbelle99} and \cite{vanbelle13}, namely: 2.7\,mas and 2.81\,mas, respectively. The intensity distribution of the DMA (Fig.~\ref{intensity_and_vis_dma}) has a radius of $\sim$20\,mas, which is comparable to the \textbf{half width half maximum (HWHM)} resulting from the GEM-FIND fit. If we adopt these numbers for the photospheric (stellar) and outer envelope radius, we can conclude that we are observing the stratification of the star out to $\sim6-7$ R$_{\star}$. Moreover, the agreement of the GEM-FIND radius with the DMA radius
supports the fact that the discrepancy between the MIDI data and the DMA are due to different flux ratios between the various shells in the photosphere and envelope, rather than a global difference in size. 

The \textbf{effective temperature} is a parameter that distinguishes the different DMAs. Literature values for RU~Vir are in the range: [$1945-2200$]~K \citep{bergeat05, lorenz01}. 
These values agree with the one derived from the Rosseland radius of the time step fitting the SED ($T_{\rm{eff}}~=~2050$~K), while the value obtained for the MIDI fit is slightly higher ($T_{\rm{eff}}~=~2545$~K).
Both temperatures are lower than the ones derived from fitting a static model to the C$_2$H$_2$ feature in the $[2.9, 3.6]~\mu$m region (3000~K) and of the static initial model of the best fitting DMAs (2800~K). As elaborated on in \cite{walter2005} dynamic effects can cause the atmospheric structure of DMAs to be fundamentally different from any static model, including the static initial model of the DMAs. Therefore, a T$_{eff}$ derived from fitting a static model atmosphere to observations is only very indirectly linked to the T$_{eff}$ of the static initial model of the best fitting DMA, thus a comparison between the two is somewhat misleading in general. In the case of the strongly pulsating atmosphere of RU~Vir, and especially for the $[2.9, 3.6]~\mu$m region we are fitting, this is particularly relevant (\citealp{loidl99}). We also note that DMAs with T$_{eff} = 3000$~K often do not produce the outflow required to reproduce the SED of RU Vir (e.g. \citealp{Erik14}). Given the highly non-static atmospheric structure of the DMA, there is no straightforward relation between these different temperatures but the results indicate that the $[2.9, 3.6]~\mu$m region is a reasonable indicator for the (static) effective temperature.

Finally, the \textbf{mass} of the best fitting dynamic model is 1.5~$M_{\odot}$. This value is in agreement with the 1.48~$M_{\odot}$ obtained from the period-mass-radius relation of \cite{vassiliadis93}.


\subsection{The shape of the SED}\label{optical_excess}

Given the success of both models in reproducing the IR-part of the SED we consider an optical excess flux more likely than an IR excess and will discuss this option first. 
The flux difference between observed photometry and the models (spectroscopy) in the $V$ band is at least a factor of $\sim~10$ in the SED. The quality of those used $BVRI$ data appears to be good as judged from later measurements, as e.g. \cite{alf12}. This author shows a lightcurve which, at the phase of the ISO spectrum, has $V \approx 11.6$. This value is even brighter than the one used in the SED (see Table~\ref{table_photometry}); moreover, the amplitude in the same band is $\approx~2$~mag, and this can still not justify our excess in $V$.

A much hotter temperature of the Mira photosphere can be ruled out as origin of the excess on the basis of the strength of the molecular features around $3$~$\mu$m. Assuming a contribution from a companion would thus be the most obvious possibility. However, the optical spectrum (i.e.\ \citealp{barnbaum92}) shows no signatures from another star and no emission lines as e.g. typical for a symbiotic system \citep{welty_wade}. 
RU~Vir has a GALEX NUV-flux but no FUV-flux \citep{galexdatarelease14}. Based on this flux and the predictions for white dwarf models \citep{bianchi11}, one can exclude the presence of a white dwarf.  One may also use the $V$ and $I$ magnitudes to assess the properties of a possible companion. However, after correcting $V$ and $I$  for the contribution from the dusty envelope as fitted with MOD, the resulting $(V-I)$ and $M_V$ cannot be reconciled with any main sequence companion. Therefore a stellar companion cannot explain the SED shape.

The presence of a sub-stellar companion that is not seen in the optical spectrum may however imply the presence of a dusty disk which would produce excess emission in the IR. A star like V~Hya would be a possible template \citep{sahai08}. The above mentioned GALEX flux is, however, a few orders of magnitude lower than what is expected for such an evolved star with an accretion disk. As evident from the MOD fit, the optical and near IR part of the SED do not show the amount of reddening as expected from the dust emission seen at longer wavelengths. Therefore any disk must not obscure the stellar photosphere, i.e.\ it should be seen almost face on. This would also be the only disk orientation compatible with the lack of evidence for asymmetries from the MIDI data.
We thus consider a disk as a very unlikely explanation. 

One aspect that has not been discussed so far is the fact that the optical photometry and the ISO spectrum were obtained many different pulsation cycles apart. While we have interpolated the optical and IR photometry to the pulsation phase of the ISO spectrum this cannot correct for any cycle-to-cycle differences. As mentioned in Sect.~\ref{phot}, RU~Vir has a long secondary period in the optical which could be an indication for such inter-cycle variations. Interestingly, also our best fitting DMA has notable inter-cycle variations. However, the optical and ISO data were obtained at similar phases of the long-term optical changes. Therefore, to explain the full SED with our best fitting DMA model, inter-cycle variations and a rather large phase uncertainty within a cycle have to be combined.

It is known, that there are differences between visual phases and DMA model phases and that these differences as well as the effects of cycle-to-cycle variations increase towards shorter wavelengths \citep{walter2}. Therefore intra-cycle variations together with phase assignment uncertainties might improve the agreement between the observed and the DMA SEDs but are probably not enough. Obtaining a broad SED within one pulsation cycle would be the only way to verify this.

\subsection{The visibilities}\label{visdiscrep} 
While the shape of the MOD visibilities are similar to the observations, the systematically lower observed level cannot be achieved by assuming a reasonable smaller distance (Sect.~\ref{hydr_mod}). Reducing the dust condensation temperature can lower the visibilities but produces the wrong shape. Therefore we believe that a simple two-component MOD model (star plus envelope with a single power law for the density) cannot reproduce the visibilities.
 
With regard to the slope of the DMA visibilities we identify two possible explanations for the discrepancies observed.

(i) The first one is model related. The models with high mass-loss rates produce a shell-like gas (and dust) component that is not observed in the data, i.e.\ the actual density distribution is smoother. This explanation is supported by the fact that models that do not produce a wind also do not exhibit the kind of slope seen in the best fit DMA. An example of the visibility of this kind of models is shown in \cite{sacuto11}. A comprehensive comparison of the dynamic model grid with interferometric observations of dusty and dust-free objects will be able to exclude or confirm this hypothesis.

(ii) Besides this  explanation, another possibility is that the environment around the star is clumpy.
This might also explain the lack of reddening mentioned in Sect.~\ref{optical_excess}. Because of the presence of clumps, the shells appear fainter than what is predicted by the model. On the other hand we do not observe any signal of clumps (i.e.\ departure from asymmetries) in the differential phases. Future observations with the second generation VLTI instrument MATISSE, will provide differential phases but also closure phases, that will be more sensitive to small asymmetries. Such kind of observations will be able to confirm the presence of clumps.

The differences between synthetic visibilities based on the DMA and MIDI visibilities around the SiC feature are most pronounced for the longest baselines, i.e.\ closest to the star. Since the formation process of SiC is not really known, the calculations assume that the SiC abundance scales with the amorphous carbon abundance as provided by the DMA. The above differences indicate that SiC could form further in and/or in a smaller amount but this has to be confirmed by comparisons based on a larger sample of stars.

The fact that the model phases best fitting the SED and the visibilities are different is probably caused by
the above mentioned cycle-to-cycle variations of the best DMA model. In the MIDI-range these variations are comparable to the intra-cycle variations, i.e.\ a certain pulsation phase within one cycle may give similar visibility profiles as a different phase in another cycle.

\subsection{The flux beyond $14~\mu$m}

The long wavelength region of the SED can be well reproduced with MOD, while the DMA always predict a too steep decline in flux towards the longest wavelengths. Tests showed that an extension of the DMA further out cannot account for this difference. Assuming a smoother density distribution of the DMA, as already argued from the visibilities and as present in the MOD model, should increase the dust emission at longer wavelengths. A more extensive comparison between DMA and a larger sample of C-rich Miras (Rau et al., in prep.) is needed to check whether the difference beyond $14~\mu$m is found only in RU~Vir or is a general characteristic of DMAs.  

A higher mass loss in the past might be an explanation although the shell parameters from MOD are close to a stationary wind solution. If the mass loss has even stopped in the recent past, this might account also for the overall differences in the shape of the SED as a very small mass loss now should shift the emission from circumstellar dust to longer wavelengths. However, besides the good fit with a stationary wind, a higher mass loss in the past would affect the visual light curve. But the average visual magnitude and the period have been stable over the last $\sim~100$ years and therefore such a scenario is also not very likely.

\section{Conclusions}\label{conclu}
As demonstrated in this study, the joint use of photometry, spectroscopy and interferometry and a comparison with models is essential to achieve a full understanding of the atmospheres of AGB stars.  

In this work we presented a study of the atmosphere of the Mira C-rich star RU~Vir. We combined spectroscopic photometric and interferometric observations, and hydrostatic and dynamic model atmospheres as well as simple geometrical models. Until now, the only studies with a similar approach as in the present study are the one by \cite{sacuto11} on R~Scl and RT~Vir \cite{sacuto13} on RT~Vir. While R~Scl is a Semi-regular pulsating star, and RT~Vir is an M-type star, RU~Vir is the first carbon-rich Mira variable for which photometry, spectroscopy \textit{and} interferometry are compared to DMAs. Studying C-rich stars is particularly important 
since DMAs are more advanced for the C-rich stars than for the O-rich stars, because the dust formation process is considered thought to be simpler and better understood in the C-rich case \citep{nowotny10,walter2}.

The HWHM derived by the GEM-FIND fitting shows overall agreement with the DMA size in the $N-$band (see Sect.~\ref{discussion}). Furthermore the Rosseland radius corresponding to the time-step best fitting the interferometric data is in agreement with the diameter derived via the $(V-K)$ relation by \cite{vanbelle99, vanbelle13}. The fitted effective temperature of the best-fitting DMA time steps is in the range of previous observational estimates but much lower than the temperature of the hydrostatic model of Sect.~\ref{MOD_fitting_procedure} and of the initial model of the best fitting DMA.

The shape of the SED in the ISO range can be well reproduced both with MOD or DMA, some discrepancies remain shortward of 2$~\mu$m and longwards of 14$~\mu$m. A similar situation exists with regard to the interferometric data, both in the shape and in the visibility level. Some of the differences might be explained by a combination of intra- and inter-cycle differences as the observations are spread over many pulsation cycles and both the star and the best fitting DMA show inter-cycle variations. Other possible reasons could be a decrease in mass loss over the last few hundred years or a sub-stellar companion associated with a dusty disk. However, these scenarios are not considered to be very likely and to check them both further observations and modelling are necessary. Based on the current work, we suspect that RU~Vir is a somewhat peculiar object. Extending the comparison between models and observations to a larger sample of C-rich Mira variables (Rau et al., in prep.) will be necessary to clarify this, and will provide the general characteristics of the atmospheres of these stars and further constraints for the models.  





\begin{acknowledgements}
The authors thank the anonymous referee for his constructive comments that helped us to improve the paper. This work is supported by the Austrian Science Fund FWF under project numbers AP$23006$, P$23586$ and P$2198$-N$16$. We are grateful to K.~Eriksson for his continuous help and valuable comments. M.~Mecina and T.~Lebzelter are thanked for many helpful discussions. This research has made use of SIMBAD database, operated at CDS, Strasbourg, France, and NASA/IPAC Infrared Science Archive. B.A. acknowledges the support from the{\em project STARKEY} funded by the ERC Consolidator Grant, G.A. n.~615604 and C.P. acknowledges financial support from the FNRS Research Fellowship – Charg\'e de Recherche.
\end{acknowledgements}

\bibliographystyle{aa} 
\bibliography{paper_hydro_ru_vir_reply.bbl}

\begin{thebibliography}{68}
\expandafter\ifx\csname natexlab\endcsname\relax\def\natexlab#1{#1}\fi

\bibitem[{{Alfonso-Garz{\'o}n} {et~al.}(2012){Alfonso-Garz{\'o}n}, {Domingo},
  {Mas-Hesse}, \& {Gim{\'e}nez}}]{alf12}
{Alfonso-Garz{\'o}n}, J., {Domingo}, A., {Mas-Hesse}, J.~M., \& {Gim{\'e}nez},
  A. 2012, \aap, 548, A79

\bibitem[{{Aringer}(2000)}]{aringer00}
{Aringer}, B. 2000, in IAU Symposium, Vol. 177, The Carbon Star Phenomenon, ed.
  R.~F. {Wing}, 519

\bibitem[{{Aringer} {et~al.}(2009){Aringer}, {Girardi}, {Nowotny}, {Marigo}, \&
  {Lederer}}]{aringer09}
{Aringer}, B., {Girardi}, L., {Nowotny}, W., {Marigo}, P., \& {Lederer}, M.~T.
  2009, \aap, 503, 913

\bibitem[{{Barnbaum}(1992)}]{barnbaum92}
{Barnbaum}, C. 1992, \apj, 385, 694

\bibitem[{{Bergeat} \& {Chevallier}(2005)}]{bergeat05}
{Bergeat}, J. \& {Chevallier}, L. 2005, \aap, 429, 235

\bibitem[{{Bianchi} {et~al.}(2014){Bianchi}, {Conti}, \&
  {Shiao}}]{galexdatarelease14}
{Bianchi}, L., {Conti}, A., \& {Shiao}, B. 2014, VizieR Online Data Catalog,
  2335, 0

\bibitem[{{Bianchi} {et~al.}(2011){Bianchi}, {Efremova}, {Herald}, {Girardi},
  {Zabot}, {Marigo}, \& {Martin}}]{bianchi11}
{Bianchi}, L., {Efremova}, B., {Herald}, J., {et~al.} 2011, \mnras, 411, 2770

\bibitem[{{Bowen}(1988)}]{bowen1988}
{Bowen}, G.~H. 1988, \apj, 329, 299

\bibitem[{{Cristallo} {et~al.}(2007){Cristallo}, {Straniero}, {Lederer}, \&
  {Aringer}}]{cristallo07}
{Cristallo}, S., {Straniero}, O., {Lederer}, M.~T., \& {Aringer}, B. 2007,
  \apj, 667, 489

\bibitem[{{Cruzal{\`e}bes} {et~al.}(2013){Cruzal{\`e}bes}, {Jorissen},
  {Rabbia}, {Sacuto}, {Chiavassa}, {Pasquato}, {Plez}, {Eriksson}, {Spang}, \&
  {Chesneau}}]{cruzal13}
{Cruzal{\`e}bes}, P., {Jorissen}, A., {Rabbia}, Y., {et~al.} 2013, \mnras, 434,
  437

\bibitem[{{Danchi} {et~al.}(1994){Danchi}, {Bester}, {Degiacomi}, {Greenhill},
  \& {Townes}}]{danchi94}
{Danchi}, W.~C., {Bester}, M., {Degiacomi}, C.~G., {Greenhill}, L.~J., \&
  {Townes}, C.~H. 1994, \aj, 107, 1469

\bibitem[{{de Graauw} {et~al.}(1996){de Graauw}, {Haser}, {Beintema},
  {Roelfsema}, {van Agthoven}, {Barl}, {Bauer}, {Bekenkamp}, {Boonstra},
  {Boxhoorn}, {Cote}, {de Groene}, {van Dijkhuizen}, {Drapatz}, {Evers},
  {Feuchtgruber}, {Frericks}, {Genzel}, {Haerendel}, {Heras}, {van der Hucht},
  {van der Hulst}, {Huygen}, {Jacobs}, {Jakob}, {Kamperman}, {Katterloher},
  {Kester}, {Kunze}, {Kussendrager}, {Lahuis}, {Lamers}, {Leech}, {van der
  Lei}, {van der Linden}, {Luinge}, {Lutz}, {Melzner}, {Morris}, {van Nguyen},
  {Ploeger}, {Price}, {Salama}, {Schaeidt}, {Sijm}, {Smoorenburg}, {Spakman},
  {Spoon}, {Steinmayer}, {Stoecker}, {Valentijn}, {Vandenbussche}, {Visser},
  {Waelkens}, {Waters}, {Wensink}, {Wesselius}, {Wiezorrek}, {Wieprecht},
  {Wijnbergen}, {Wildeman}, \& {Young}}]{isosws}
{de Graauw}, T., {Haser}, L.~N., {Beintema}, D.~A., {et~al.} 1996, \aap, 315,
  L49

\bibitem[{{Eggen}(1975)}]{eggen1975a}
{Eggen}, O.~J. 1975, \apjs, 29, 77

\bibitem[{{Eriksson} {et~al.}(2014){Eriksson}, {Nowotny}, {H{\"o}fner},
  {Aringer}, \& {Wachter}}]{Erik14}
{Eriksson}, K., {Nowotny}, W., {H{\"o}fner}, S., {Aringer}, B., \& {Wachter},
  A. 2014, \aap, 566, A95

\bibitem[{{Fleischer} {et~al.}(1992){Fleischer}, {Gauger}, \&
  {Sedlmayr}}]{Fleischer92}
{Fleischer}, A.~J., {Gauger}, A., \& {Sedlmayr}, E. 1992, \aap, 266, 321

\bibitem[{{Gail} \& {Sedlmayr}(1988)}]{gail88}
{Gail}, H.-P. \& {Sedlmayr}, E. 1988, \aap, 206, 153

\bibitem[{{Gail} \& {Sedlmayr}(2013)}]{gail}
{Gail}, H.-P. \& {Sedlmayr}, E. 2013, {Physics and Chemistry of Circumstellar
  Dust Shells}

\bibitem[{{Gauger} {et~al.}(1990){Gauger}, {Sedlmayr}, \& {Gail}}]{gauger90}
{Gauger}, A., {Sedlmayr}, E., \& {Gail}, H.-P. 1990, \aap, 235, 345

\bibitem[{{Gautschy-Loidl} {et~al.}(2004){Gautschy-Loidl}, {H{\"o}fner},
  {J{\o}rgensen}, \& {Hron}}]{Gautschy-Loidl04}
{Gautschy-Loidl}, R., {H{\"o}fner}, S., {J{\o}rgensen}, U.~G., \& {Hron}, J.
  2004, \aap, 422, 289

\bibitem[{{Groenewegen}(2012)}]{mod}
{Groenewegen}, M.~A.~T. 2012, \aap, 543, A36

\bibitem[{{H{\"o}fner}(1999)}]{hofner99}
{H{\"o}fner}, S. 1999, \aap, 346, L9

\bibitem[{{H{\"o}fner}(2007)}]{hofner07}
{H{\"o}fner}, S. 2007, in Astronomical Society of the Pacific Conference
  Series, Vol. 378, Why Galaxies Care About AGB Stars: Their Importance as
  Actors and Probes, ed. F.~{Kerschbaum}, C.~{Charbonnel}, \& R.~F. {Wing}, 145

\bibitem[{{H{\"o}fner} \& {Dorfi}(1997)}]{HoefnerDorfi}
{H{\"o}fner}, S. \& {Dorfi}, E.~A. 1997, \aap, 319, 648

\bibitem[{{H{\"o}fner} {et~al.}(2003){H{\"o}fner}, {Gautschy-Loidl}, {Aringer},
  \& {J{\o}rgensen}}]{hofner03}
{H{\"o}fner}, S., {Gautschy-Loidl}, R., {Aringer}, B., \& {J{\o}rgensen}, U.~G.
  2003, \aap, 399, 589

\bibitem[{{Hron} {et~al.}(1997){Hron}, {Loidl}, \& {Kerschbaum}}]{hron98}
{Hron}, J., {Loidl}, R., \& {Kerschbaum}, F. 1997, \apss, 251, 211

\bibitem[{{Iben} \& {Renzini}(1983)}]{ibenrenzini83}
{Iben}, Jr., I. \& {Renzini}, A. 1983, \araa, 21, 271

\bibitem[{{Ivezi\'c} \& {Elitzur}(1995)}]{dustypaper}
{Ivezi\'c}, Z. \& {Elitzur}, M. 1995, \apj, 445, 415

\bibitem[{{Ivezi\'c} {et~al.}(1997){Ivezi\'c}, {Groenewegen}, {Men'shchikov},
  \& {Szczerba}}]{ivezic97}
{Ivezi\'c}, Z., {Groenewegen}, M.~A.~T., {Men'shchikov}, A., \& {Szczerba}, R.
  1997, \mnras, 291, 121

\bibitem[{{Jaffe}(2004)}]{jaffe04}
{Jaffe}, W.~J. 2004, in Society of Photo-Optical Instrumentation Engineers
  (SPIE) Conference Series, Vol. 5491, New Frontiers in Stellar Interferometry,
  ed. W.~A. {Traub}, 715

\bibitem[{{J{\o}rgensen} {et~al.}(2000){J{\o}rgensen}, {Hron}, \&
  {Loidl}}]{jorgensen00}
{J{\o}rgensen}, U.~G., {Hron}, J., \& {Loidl}, R. 2000, \aap, 356, 253

\bibitem[{{Klotz} {et~al.}(2013){Klotz}, {Paladini}, {Hron}, {Aringer},
  {Sacuto}, {Marigo}, \& {Verhoelst}}]{txpsc}
{Klotz}, D., {Paladini}, C., {Hron}, J., {et~al.} 2013, \aap, 550, A86

\bibitem[{{Klotz} {et~al.}(2012{\natexlab{a}}){Klotz}, {Sacuto}, {Kerschbaum},
  {Paladini}, {Olofsson}, \& {Hron}}]{klotz12}
{Klotz}, D., {Sacuto}, S., {Kerschbaum}, F., {et~al.} 2012{\natexlab{a}}, \aap,
  541, A164

\bibitem[{{Klotz} {et~al.}(2012{\natexlab{b}}){Klotz}, {Sacuto}, {Paladini},
  {Hron}, \& {Wachter}}]{gemfind}
{Klotz}, D., {Sacuto}, S., {Paladini}, C., {Hron}, J., \& {Wachter}, G.
  2012{\natexlab{b}}, in Society of Photo-Optical Instrumentation Engineers
  (SPIE) Conference Series, Vol. 8445, Society of Photo-Optical Instrumentation
  Engineers (SPIE) Conference Series, 1

\bibitem[{{Leinert} {et~al.}(2003){Leinert}, {Graser}, {Richichi},
  {Sch{\"o}ller}, {Waters}, {Perrin}, {Jaffe}, {Lopez}, {Glazenborg-Kluttig},
  {Przygodda}, {Morel}, {Biereichel}, {Haddad}, {Housen}, \&
  {Wallander}}]{midi2003}
{Leinert}, C., {Graser}, U., {Richichi}, A., {et~al.} 2003, The Messenger, 112,
  13

\bibitem[{{Lobel} {et~al.}(1999){Lobel}, {Doyle}, \& {Bagnulo}}]{lobel99}
{Lobel}, A., {Doyle}, J.~G., \& {Bagnulo}, S. 1999, \aap, 343, 466

\bibitem[{{Loidl} {et~al.}(1999){Loidl}, {H{\"o}fner}, {J{\o}rgensen}, \&
  {Aringer}}]{loidl99}
{Loidl}, R., {H{\"o}fner}, S., {J{\o}rgensen}, U.~G., \& {Aringer}, B. 1999,
  \aap, 342, 531

\bibitem[{{Loidl} {et~al.}(2001){Loidl}, {Lan{\c c}on}, \&
  {J{\o}rgensen}}]{loidl01}
{Loidl}, R., {Lan{\c c}on}, A., \& {J{\o}rgensen}, U.~G. 2001, \aap, 371, 1065

\bibitem[{{Lorenz-Martins} {et~al.}(2001){Lorenz-Martins}, {de Ara{\'u}jo},
  {Codina Landaberry}, {de Almeida}, \& {de Nader}}]{lorenz01}
{Lorenz-Martins}, S., {de Ara{\'u}jo}, F.~X., {Codina Landaberry}, S.~J., {de
  Almeida}, W.~G., \& {de Nader}, R.~V. 2001, \aap, 367, 189

\bibitem[{{Mart{\'{\i}}-Vidal} {et~al.}(2011){Mart{\'{\i}}-Vidal}, {Marcaide},
  {Quirrenbach}, {Ohnaka}, {Guirado}, \& {Wittkowski}}]{martividal11}
{Mart{\'{\i}}-Vidal}, I., {Marcaide}, J.~M., {Quirrenbach}, A., {et~al.} 2011,
  \aap, 529, A115

\bibitem[{{Mattsson} \& {H{\"o}fner}(2011)}]{mattsson11}
{Mattsson}, L. \& {H{\"o}fner}, S. 2011, \aap, 533, A42

\bibitem[{{Mattsson} {et~al.}(2010){Mattsson}, {Wahlin}, \&
  {H{\"o}fner}}]{mattsson10}
{Mattsson}, L., {Wahlin}, R., \& {H{\"o}fner}, S. 2010, \aap, 509, A14

\bibitem[{{Neilson} \& {Lester}(2008)}]{neilson08}
{Neilson}, H.~R. \& {Lester}, J.~B. 2008, \aap, 490, 807

\bibitem[{{Nowotny} {et~al.}(2011){Nowotny}, {Aringer}, {H{\"o}fner}, \&
  {Lederer}}]{walter2}
{Nowotny}, W., {Aringer}, B., {H{\"o}fner}, S., \& {Lederer}, M.~T. 2011, \aap,
  529, A129

\bibitem[{{Nowotny} {et~al.}(2010){Nowotny}, {H{\"o}fner}, \&
  {Aringer}}]{nowotny10}
{Nowotny}, W., {H{\"o}fner}, S., \& {Aringer}, B. 2010, \aap, 514, A35

\bibitem[{{Nowotny} {et~al.}(2005){Nowotny}, {Lebzelter}, {Hron}, \&
  {H{\"o}fner}}]{walter2005}
{Nowotny}, W., {Lebzelter}, T., {Hron}, J., \& {H{\"o}fner}, S. 2005, \aap,
  437, 285

\bibitem[{{Ohnaka} {et~al.}(2007){Ohnaka}, {Driebe}, {Weigelt}, \&
  {Wittkowski}}]{ohnaka07}
{Ohnaka}, K., {Driebe}, T., {Weigelt}, G., \& {Wittkowski}, M. 2007, \aap, 466,
  1099

\bibitem[{{Paladini} {et~al.}(2009){Paladini}, {Aringer}, {Hron}, {Nowotny},
  {Sacuto}, \& {H{\"o}fner}}]{paladini09}
{Paladini}, C., {Aringer}, B., {Hron}, J., {et~al.} 2009, \aap, 501, 1073

\bibitem[{{Paladini} {et~al.}(2011){Paladini}, {van Belle}, {Aringer}, {Hron},
  {Reegen}, {Davis}, \& {Lebzelter}}]{paladini11}
{Paladini}, C., {van Belle}, G.~T., {Aringer}, B., {et~al.} 2011, \aap, 533,
  A27

\bibitem[{{Pegourie}(1988)}]{pegourie}
{Pegourie}, B. 1988, \aap, 194, 335

\bibitem[{{Percy} \& {Bagby}(1999)}]{percy99}
{Percy}, J.~R. \& {Bagby}, D.~H. 1999, \pasp, 111, 203

\bibitem[{{Pitman} {et~al.}(2008){Pitman}, {Hofmeister}, {Corman}, \&
  {Speck}}]{pitman08}
{Pitman}, K.~M., {Hofmeister}, A.~M., {Corman}, A.~B., \& {Speck}, A.~K. 2008,
  \aap, 483, 661

\bibitem[{{Rouleau} \& {Martin}(1991)}]{roleau_and_martin}
{Rouleau}, F. \& {Martin}, P.~G. 1991, \apj, 377, 526

\bibitem[{{Sacuto} {et~al.}(2011){Sacuto}, {Aringer}, {Hron}, {Nowotny},
  {Paladini}, {Verhoelst}, \& {H{\"o}fner}}]{sacuto11}
{Sacuto}, S., {Aringer}, B., {Hron}, J., {et~al.} 2011, \aap, 525, A42

\bibitem[{{Sacuto} {et~al.}(2013){Sacuto}, {Ramstedt}, {H{\"o}fner},
  {Olofsson}, {Bladh}, {Eriksson}, {Aringer}, {Klotz}, \&
  {Maercker}}]{sacuto13}
{Sacuto}, S., {Ramstedt}, S., {H{\"o}fner}, S., {et~al.} 2013, \aap, 551, A72

\bibitem[{{Sahai} {et~al.}(2008){Sahai}, {Findeisen}, {Gil de Paz}, \&
  {S{\'a}nchez Contreras}}]{sahai08}
{Sahai}, R., {Findeisen}, K., {Gil de Paz}, A., \& {S{\'a}nchez Contreras}, C.
  2008, \apj, 689, 1274

\bibitem[{{Samus} {et~al.}(2009){Samus}, {Kazarovets}, {Pastukhova},
  {Tsvetkova}, \& {Durlevich}}]{GCVS}
{Samus}, N.~N., {Kazarovets}, E.~V., {Pastukhova}, E.~N., {Tsvetkova}, T.~M.,
  \& {Durlevich}, O.~V. 2009, \pasp, 121, 1378

\bibitem[{{Sloan} {et~al.}(2003){Sloan}, {Kraemer}, \& {Price}}]{sloan03}
{Sloan}, G.~C., {Kraemer}, K.~E., \& {Price}, S.~D. 2003, in ESA Special
  Publication, Vol. 481, The Calibration Legacy of the ISO Mission, ed.
  L.~{Metcalfe}, A.~{Salama}, S.~B. {Peschke}, \& M.~F. {Kessler}, 447

\bibitem[{{van Belle}(1999)}]{vanbelle99}
{van Belle}, G.~T. 1999, \pasp, 111, 1515

\bibitem[{{van Belle} {et~al.}(2013){van Belle}, {Paladini}, {Aringer}, {Hron},
  \& {Ciardi}}]{vanbelle13}
{van Belle}, G.~T., {Paladini}, C., {Aringer}, B., {Hron}, J., \& {Ciardi}, D.
  2013, \apj, 775, 45

\bibitem[{{Vassiliadis} \& {Wood}(1993)}]{vassiliadis93}
{Vassiliadis}, E. \& {Wood}, P.~R. 1993, \apj, 413, 641

\bibitem[{{Welty} \& {Wade}(1995)}]{welty_wade}
{Welty}, A.~D. \& {Wade}, R.~A. 1995, \aj, 109, 326

\bibitem[{{Whitelock} {et~al.}(2006){Whitelock}, {Feast}, {Marang}, \&
  {Groenewegen}}]{whitelock06}
{Whitelock}, P.~A., {Feast}, M.~W., {Marang}, F., \& {Groenewegen}, M.~A.~T.
  2006, \mnras, 369, 751

\bibitem[{{Winters} {et~al.}(1995){Winters}, {Fleischer}, {Gauger}, \&
  {Sedlmayr}}]{winters95}
{Winters}, J.~M., {Fleischer}, A.~J., {Gauger}, A., \& {Sedlmayr}, E. 1995,
  \aap, 302, 483

\bibitem[{{Wittkowski} {et~al.}(2008){Wittkowski}, {Boboltz}, {Driebe}, {Le
  Bouquin}, {Millour}, {Ohnaka}, \& {Scholz}}]{witt08}
{Wittkowski}, M., {Boboltz}, D.~A., {Driebe}, T., {et~al.} 2008, \aap, 479, L21

\bibitem[{{Wittkowski} {et~al.}(2011){Wittkowski}, {Boboltz}, {Ireland},
  {Karovicova}, {Ohnaka}, {Scholz}, {van Wyk}, {Whitelock}, {Wood}, \&
  {Zijlstra}}]{witt11}
{Wittkowski}, M., {Boboltz}, D.~A., {Ireland}, M., {et~al.} 2011, \aap, 532, L7

\bibitem[{{Wittkowski} {et~al.}(2001){Wittkowski}, {Hummel}, {Johnston},
  {Mozurkewich}, {Hajian}, \& {White}}]{witt01}
{Wittkowski}, M., {Hummel}, C.~A., {Johnston}, K.~J., {et~al.} 2001, \aap, 377,
  981

\bibitem[{{Woitke}(2003)}]{woitke03}
{Woitke}, P. 2003, in IAU Symposium, Vol. 210, Modelling of Stellar
  Atmospheres, ed. N.~{Piskunov}, W.~W. {Weiss}, \& D.~F. {Gray}, 387

\bibitem[{{Yamamura} \& {de Jong}(2000)}]{yam00}
{Yamamura}, I. \& {de Jong}, T. 2000, in ESA Special Publication, Vol. 456, ISO
  Beyond the Peaks: The 2nd ISO Workshop on Analytical Spectroscopy, ed.
  A.~{Salama}, M.~F. {Kessler}, K.~{Leech}, \& B.~{Schulz}, 155

\end{thebibliography}

\Online
\begin{appendix} 

\section{Observing log}

\begin{table*}
\caption{\label{tab_midi_observ} Journal of the MIDI observations of RU~Vir.} 
\centering
\begin{tabular}{llllllllll}
\hline\hline
Target & UT date \& time & Config. & $B_p$ & PA & Seeing & Airmass &   Mode  & $\phi$\\
& & & [m] & [$^\circ$]&[''] &\\
\hline\hline
RU~Vir & 2010-05-23 T01:23:12 	&  	 H0-E0 	& 47.2 &
	 73 	& 1.14 &
 	1.52	 &     SCI-PHOT  & 0.11 \\

HD 120323 & 2010-05-23 T00:32:16 &
	\ldots &
	45.6 &	53 &
	1.14 &	1.12 &   \ldots & \ldots \\

 HD 81797  &
	2010-05-23 T23:05:53 &
	\ldots &	47.4 &	73  &	0.99 &
	1.06 &    \ldots   & \ldots \\
 \ldots &	2010-05-23 T00:00:32  &
	\ldots &	44.6 &	74 &
	0.70 &
	1.18 &    \ldots & \ldots\\
 HD 133216  &
	2010-05-23 T02:18:36 &
	\ldots &	65.0 &	98 & 1.14	 &
	1.05 &    \ldots  & \ldots\\	

\hline
RU~Vir &	 2010-06-04 T01:30:24 	&  H0-E0	 &	47.9 &
	72   &   1.45 &	1.17 &    \ldots & 0.14\\
HD 120323 &
	2010-06-04 T02:06:52 &
	\ldots &
	43.8 &	45 &
	1.43 &	1.21 &   \ldots &   \ldots \\
\ldots &	2010-06-04 T00:50:11  &
	\ldots &	47.6 & 63 & 1.41
	 &  1.03
	 &   \ldots &   \ldots \\
\ldots  &
	2010-06-04 T03:03:39 &
	\ldots &	45.2 &	82 &	 1.20 &
	1.07 &    \ldots   &   \ldots \\
\ldots &
	2010-06-04 T01:12:54 &
	\ldots &	47.9 & 67 &	1.63 &
	1.02 &      \ldots  &   \ldots \\
\ldots &
	2010-06-04 T01:50:48 &
	 \ldots &	47.8 &	72 &	 1.53 &
	1.02 &      \ldots &   \ldots \\
HD 133216  &
	2010-06-04 T03:46:05 &
	\ldots &	47.0 &	68 &	1.43 &
	1.17 &    \ldots &   \ldots \\	
\hline
\hline
RU~Vir &	 2014-04-11 T00:48:17 	&  D0-G1	 &	67.6 &
	134   & 0.96 &	1.52 &    HIGH-SENS  &   0.37 \\
HD 120323 &
	2014-04-11 T01:03:12 &
	\ldots &
	45.3 &	112 &
	0.79 &	1.70 &  \ldots &  \ldots  \\
\ldots &	2014-04-11 T01:28:11  &
	\ldots &	 49.7 
	& 113	&  0.96 &
	1.52 & \ldots &   \ldots \\	
\hline	
RU~Vir &	 2014-04-11 T01:15:18 	& \ldots 	 &	69.4 &
	132   & 1.00 &	1.59 &  \ldots  &  0.37   \\	
HD 120323 &
	2014-04-11 T01:03:12 &
	\ldots &
	45.3 &	112 &
	0.79 &	1.70 &  \ldots &   \ldots \\
\ldots &	2014-04-11 T01:28:11  &
	\ldots &	 49.7 
	& 113	&  0.96 &
	1.52 & \ldots &   \ldots \\	
\hline	
RU~Vir &	 2014-04-11 T02:11:25 	&  H0-I1	 &	40.5 &
	141   & 1.02 &	1.32 &  \ldots  &   0.37  \\	
HD 120323  &
	2014-04-11 T02:25:17 &
	\ldots &	35.2 & 134 &	 0.95 &
	1.26 &  \ldots  &   \ldots \\	
HD  81797 &
	2014-04-11 T02:55:16 &
	\ldots &	36.5 &	163 &	 1.03 &
	1.21 &  \ldots &   \ldots \\
  \ldots &
	2014-04-11 T03:20:45 &
	\ldots &	36.0 &	168 &	1.13 &
	1.30 &  \ldots &   \ldots \\		
\hline	
RU~Vir &	 2014-04-11 T02:38:08 	&  H0-I1	 &	40.0 &
	142   & 0.98 &	1.24 &  \ldots &   0.37 \\	
HD 120323	  &
	2014-04-11 T01:58:26 &
	\ldots &	33.7 &	133 &	 0.80 &
	1.36 &  \ldots  &   \ldots \\
\ldots  &
	2014-04-11 T02:25:17 &
	\ldots &	35.2 & 134 &	 0.95 &
	1.26 &  \ldots  &   \ldots \\	
HD  81797 &
	2014-04-11 T02:55:16 &
	\ldots &	36.5 &	163 &	 1.03 &
	1.21 &   \ldots &   \ldots \\
\ldots &
	2014-04-11 T03:20:45 &
	\ldots &	36.0 &	168 &	1.13 &
	1.30 &  \ldots &   \ldots \\
\hline	
RU~Vir &	 2014-04-11 T03:07:40 	&  H0-I1	 &	39.3 &
	143   & 1.05 &	1.18 &  \ldots &   0.37  \\
HD 120323  &
	2014-04-11 T02:25:17 &
\ldots &	35.2 & 134 &	 0.95 &
	1.26 &  \ldots   &   \ldots  \\	
HD  81797 &
	2014-04-11 T02:55:16 &
	\ldots &	36.5 &	163 &	 1.03 &
	1.21 &   \ldots  &   \ldots \\
\ldots &
	2014-04-11 T03:20:45 &
	\ldots &	36.0 &	168 &	1.13 &
	1.30 &  \ldots  &   \ldots \\
\hline	
RU~Vir &	 2014-04-11 T03:58:15 	&  D0-G1	 &	68.4 &
	130   & 1.18 &	1.14 &  \ldots   &   0.37  \\	
HD 120323  &
	2014-04-11 T03:45:37 &
	\ldots &	 66.7 &	124 &	 1.32 &
	1.07 &  \ldots &   \ldots \\
\ldots &
	2014-04-11 T04:11:45 &
	\ldots &	68.5 &	127 &	1.20 &
	1.04 & \ldots &   \ldots \\
\ldots  &
	2014-04-11 T04:39:44 &
	\ldots &	69.9 &	130 &	 1.38 &
	1.02 &  \ldots  &   \ldots \\
\ldots  &
	2014-04-11 T05:47:28 &
	\ldots &	71.5 & 138 &	1.12 &
	1.03 &   \ldots  &   \ldots \\
\hline
	
RU~Vir &	 2014-04-11 T04:25:14	&  D0-G1		 &	66.2 &
	131   & 1.28 &	1.14 &   \ldots &   0.37  \\	
HD 120323  &
	2014-04-11 T03:45:37 &
	\ldots &	 66.7 &	124 &	 1.32 &
	1.07 &  \ldots   &   \ldots \\
\ldots &
	2014-04-11 T04:11:45 &
	\ldots &	68.5 &	127 &	1.20 &
	1.04 &  \ldots &   \ldots \\
\ldots	  &
	2014-04-11 T04:39:44 &
	\ldots &	69.9 &	130 &	 1.38 &
	1.02 & \ldots &   \ldots \\
\ldots &
	2014-04-11 T05:47:28 &
	\ldots &	71.5 & 138 &	1.12 &
	1.03 &  \ldots   &   \ldots \\
\hline	
RU~Vir &	 2014-04-11 T04:53:53 	& 	D0-G1	 &	63.4 &
	133   & 1.26 &	1.16 &  \ldots  &   0.37  \\
HD 120323  &
	2014-04-11 T03:45:37 &
	\ldots &	 66.7 &	124 &	 1.32 &
	1.07 &   \ldots &   \ldots \\
\ldots &
	2014-04-11 T04:11:45 &
	\ldots &	68.5 &	127 &	1.20 &
	1.04 &  \ldots &   \ldots \\
\ldots  &
	2014-04-11 T04:39:44 &
	 \ldots &	69.9 &	130 &	 1.38 &
	1.02 &  \ldots  &   \ldots \\
\ldots  &
	2014-04-11 T05:47:28 &
	\ldots &	71.5 & 138 &	1.12 &
	1.03 &  \ldots   &   \ldots \\
\hline	
RU~Vir &	 2014-04-11 T05:19:21 	& D0-G1	&	60.6 &
	135   & 1.10 &	1.20 &  \ldots &   0.37  \\	

HD  120323 &
	2014-04-11 T04:11:45 &
	\ldots &	68.5 &	127 &	1.20 &
	1.04 &  \ldots  &   \ldots \\
\ldots  &
	2014-04-11 T05:47:28 &
	 \ldots &	 71.5 & 138 &	1.12 &
	1.03 &    \ldots &   \ldots \\
\hline	
RU~Vir &	 2014-04-11 T05:59:56 	&  D0-G1	 &	55.6 &
	139   & 1.09 &	1.30 &  \ldots &   0.37  \\
HD  120323 &
	2014-04-11 T04:11:45 &
	\ldots &	68.5 &	127 &	1.20 &
	1.04 &  \ldots  &   \ldots \\
\ldots  &
	2014-04-11 T05:47:28 &
	\ldots &	71.5 & 138 &	1.12 &
	1.03 &   \ldots   &   \ldots \\		
\hline	
\end{tabular}
\end{table*}

    \begin{figure*}
   \centering
   \includegraphics[width=\hsize, bb=66 364 552 718]{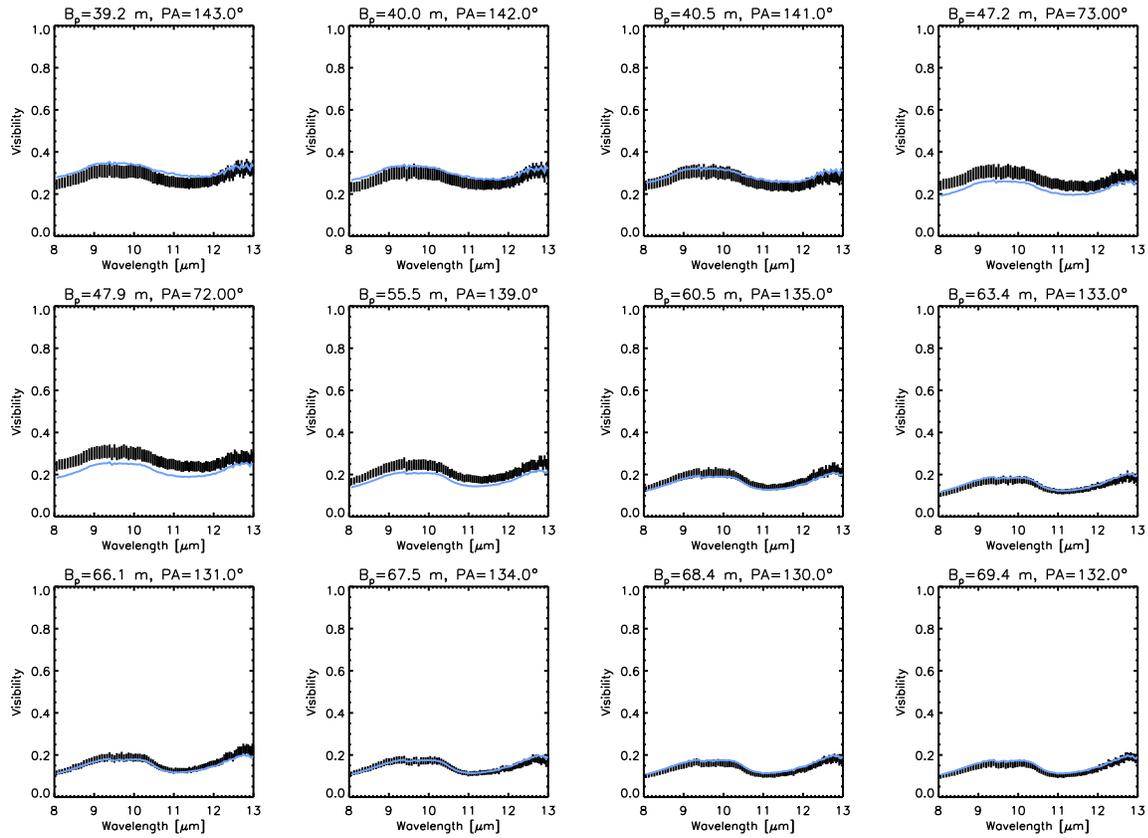}
      \caption{Comparison of the observational MIDI data with the best-fitting geometrical model (See Table~\ref{table_gemfind_results}). Shown are visibilities vs. wavelength at different position angles and baseline lengths (See Fig.~\ref{ruvir_calibr_vis} for the complete set of visibilities). Black lines represent the MIDI observations of RU~Vir. Light blue lines represent the fit of the best-fitting UD+Gaussian model.}
         \label{fit_gemfind_wave-vis__complete}
   \end{figure*}

\end{appendix}

\end{document}